\documentclass[11pt,a4paper]{article}
\pdfoutput=1
\usepackage{jheppub}
\usepackage[utf8]{inputenc}

\usepackage{color}
\usepackage[dvipsnames]{xcolor}
\usepackage{soul}
\usepackage{amsmath}
\usepackage{amssymb}
\usepackage{mathtools}
\usepackage{bbold}
\usepackage{caption}
\usepackage{amsfonts}
\usepackage{graphicx}
\usepackage{multirow}
\usepackage{slashed}
\usepackage{cancel}
\usepackage{cleveref}
\usepackage[normalem]{ulem}
\usepackage{verbatim}

\usepackage{mathtools}

\definecolor{flag}{rgb}{1.0, 0.03, 0.0}

\newcommand{\be}{\begin{equation}}
\newcommand{\ee}{\end{equation}}
\newcommand{\bea}{\begin{align}}
\newcommand{\eea}{\end{align}}

\newcommand{\ft}{\tilde{f}}

\newcommand{\w}{\text{w}}
\newcommand{\oned}{\text{1}}
\newcommand{\eq}{\text{eq}}

\newcommand{\gstars}{g_{*\text{\tiny{\it{S}}}}}

\newcommand{\meter}{{\, {\rm m}}}

\newcommand{\cm}{{\, {\rm cm}}}

\newcommand{\eV}{{\, {\rm eV}}}
\newcommand{\keV}{{\, {\rm keV}}}
\newcommand{\MeV}{{\, {\rm MeV}}}
\newcommand{\GeV}{{\, {\rm GeV}}}

\newcommand{\Tesla}{{\, {\rm T}}}

\title{Axion Interactions with Domain and Bubble Walls}

\author[a]{Isabel Garcia Garcia}
\emailAdd{isabelgg@uw.edu}
\author[b]{and Rudin Petrossian-Byrne}
\emailAdd{rpetross@ictp.it}

\affiliation[a]{Department of Physics, University of Washington, Seattle, WA 98195, USA}
\affiliation[b]{Abdus Salam International Centre for Theoretical Physics,
Strada Costiera 11, 34151, Trieste, Italy}

\abstract{
We show that interactions between axion-like particles (ALPs) and co-dimension one defects, such as phase-transition bubble walls and solitonic domain walls, can lead to important changes in the evolution of both walls and ALPs. The leading effect arises from the change in the ALP decay constant across the interface, which naturally follows from shift-symmetric interactions with the corresponding order parameter. Specifically, we show that for thin walls moving relativistically, an ALP background -- such as e.g. axion dark matter -- gives rise to a frictional force on the interface that is proportional to $\gamma^2$, with $\gamma$ the Lorentz factor of the wall, and that this effect is present in both the oscillating and frozen axion regimes. We explore the broader consequences of this effect for bubble and domain walls in the early universe, and show that this source of friction can be present even in the absent of a conventional medium such as radiation or matter. Possible implications include modifications to the dynamics of bubble and domain walls and their corresponding gravitational wave signatures, as well as the generation of a dark radiation component of ALPs in the form of ultra-relativistic `axion shells' with Lorentz factor $\gamma_\text{shell} \simeq 2\gamma^2 \gg 1$ that may remain relativistic until the present day.
}

\begin{document}

\maketitle

\newpage 

\section{Introduction}
\label{sec:intro}

Symmetry-restoration at high temperature is a generic phenomenon in quantum field theory. In the early universe, this leads to the expectation that spontaneous symmetry breaking occurred as the universe cooled down, most notably at electroweak and QCD scale temperatures. Lattice data suggests that both processes took place adiabatically, in the form of a cross-over between the symmetry-restored and symmetry-broken phases~\cite{Bonati:2013tqa,Borsanyi:2016ksw,Kajantie:1996mn,Kajantie:1996qd,Laine:1998jb,Csikor:1998eu}. Moreover, the Standard Model (SM) lacks the existence of (even unstable) non-perturbative defects that could have been formed as a result of spontaneous symmetry breaking. In contrast, in extensions of the SM spontaneous symmetry breaking is often accompanied by the production of soliton-like defects such as cosmic strings, domain walls, or the vacuum bubbles of a first order phase transition. These extended objects are common in the non-perturbative spectrum of many theories beyond the SM. Inaccessible in the lab, they can only be produced in the early Universe.

Understanding the phenomenology of these objects is essential to probe the phase structure of the Universe at high temperatures and to explore physics beyond the SM in the non-perturbative regime. A crucial ingredient concerns interactions among these extended objects and perturbative degrees of freedom (dof). In the early universe, this will include the primordial plasma as well as other particle species not in thermal equilibrium with the SM, such as e.g.~the dark matter (DM). A notable example of the importance of interactions between particles and defects includes axion production from cosmic strings, which constitutes one of the main production mechanisms for axion dark matter. A second example concerns the friction on expanding bubble walls during a first order phase transition due to interactions with the thermal fluid. This source of pressure can preclude the acceleration of the bubble walls, and qualitatively alter experimental signatures, such as the strength and shape of the resulting gravitational wave (GW) signal (see e.g.~\cite{Caprini:2015zlo,Caprini:2019egz} for reviews).

In this work, we discuss a new set of physical effects arising from the interactions between co-dimension one wall-like defects -- whether the (meta)stable domain walls of a spontaneously broken discrete symmetry or the bubble walls of a phase transition -- and axion-like particles (ALPs). ALPs are a common occurrence both in the SM (pions) and its UV-completions. The Peccei-Quinn solution to the strong CP problem provides strong motivation for the existence of the QCD axion \cite{Peccei:1977hh,Weinberg:1977ma,Wilczek:1977pj}, and in the context of string theory axion-like particles are believed to be ubiquitous~\cite{Arvanitaki:2009fg}. In this work, we will loosely use the term `ALP" or `axion" to refer to any pseudo-scalar particle with an approximate shift-symmetry. Shift-symmetry preserving interactions are suppressed by the axion decay constant, and breaking of the shift-symmetry generically leads to a non-trivial potential for the axion field. Moreover, if axions exist, they are naturally produced in the early universe, either through the dynamics of a post-inflationary network of axion strings~\cite{Sikivie:1982qv,Vilenkin:1982ks,Vilenkin:1984ib,Davis:1986xc,Gorghetto:2018myk} or via the misalignment mechanism~\cite{Preskill:1982cy,Abbott:1982af,Dine:1982ah}. Barring fine-tuning, a contribution to the total energy budget in the form of axions is therefore a general expectation in the early universe and -- if the axion is sufficiently long-lived -- at the current epoch.

In general, shift-symmetry preserving interactions between axions and the scalar order parameter characterizing spontaneous symmetry breaking will form part of the axion effective theory. In particular, the following operator involving the (dimensionless) axion $\theta$ and any scalar field $\phi$ that is part of the axion EFT cannot be forbidden on the basis of symmetry:
\be \label{eq:dim6}
        \Delta \mathcal{L}_\text{EFT}  = \frac{1}{2} \kappa \phi^2 (\partial_\mu \theta)^2 \ ,
\ee
with $\kappa$ a dimensionless coupling. We emphasize that $\phi$ may be an elementary scalar or an effective composite field characterizing e.g.~some non-perturbative condensate of the UV theory, and it may in general be real or complex (if the latter, then $\phi \rightarrow |\phi|$ in Eq.\eqref{eq:dim6}).~\footnote{An even lower dimensional interaction corresponds to the operator $\phi (\partial_\mu \theta)^2 $. Qualitatively, the effect of this interaction is equivalent to that in Eq.\eqref{eq:dim6}, and an operator linear in $\phi$ could be absent if e.g.~$\phi$ was charged under a $\mathbb{Z}_2$-symmetry. We will therefore focus on Eq.\eqref{eq:dim6} except where indicated explicitly.} If $\phi$ acquires a non-zero vacuum expectation value (vev) $\langle \phi \rangle = v$ this interaction will lead to a change in the axion decay constant by an amount $\Delta f^2 = \kappa v^2$. If spontaneous symmetry breaking is accompanied by the formation of defects, such as vacuum bubbles or domain walls, axions will feature non-trivial interactions with these objects as a result of Eq.\eqref{eq:dim6}. These interactions, and their effect on the evolution of the corresponding defects in the early universe, are the main focus on this work.

\begin{figure}
    \centering
    \includegraphics[scale=0.95]{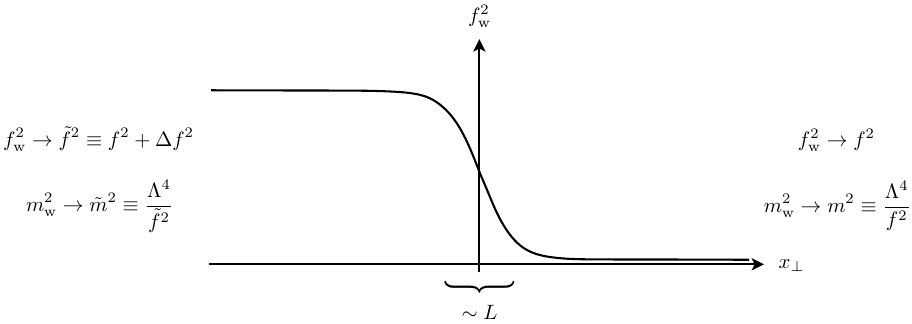}
    \caption{In general, the decay constant of an axion can change across a domain or bubble wall formed in the early universe as a result of spontaneous symmetry breaking. The change takes place smoothly over some distance $L$ corresponding to the wall thickness. Typically, a change in decay constant also leads to a change in the mass of the canonically normalized axion field, as indicated in the figure. WLOG, we take the decay constant to vary from $f^2$ to the right of the wall to $\tilde f^2 = f^2 + \Delta f^2$ to the left of the wall.} 
    \label{fig:setup}
\end{figure}

In the presence of a coupling of the general form Eq.\eqref{eq:dim6}, axion-wall interactions are then well-described, to leading order in a derivative expansion, in terms of an axion field on the background of a space-time varying decay constant, i.e.~
\be
    f_\text{w}^2 (x) = f^2 + \kappa v^2(x) \ .
\ee
In a realistic situation, $f_\w^2$ will vary smoothly across the interface over some distance corresponding to the wall thickness. In general, shift-symmetry breaking effects will generate an axion mass of the form $m^2 = \Lambda^4 / f^2$, with $\Lambda$ the scale of shift-symmetry breaking that is independent of $f$. As a result, a change in $f$ generally also leads to a change in mass. As we will see, the effect of a change in mass in the evolution of the axion-wall system is subleading compared to that of a change in decay constant. This basic setup is schematically illustrated in Fig.~\ref{fig:setup}.

It will be important in the following that walls can be characterized as `thin' or `thick' depending on whether the wall thickness \emph{in the wall rest frame}, $L$, is either $L \ll (\gamma m)^{-1}$ or $L \gg (\gamma m)^{-1}$ respectively. Intuitively, a wall is considered `thin' if the axion de Broglie wavelength in the wall frame exceeds the wall thickness. Crucially, this criterion depends on the velocity of the wall, and any wall can eventually become `thick' for sufficiently large $\gamma$-factors. (Notice this is different from the usual terminology used to refer to walls as thin or thick depending on how their thickness compares to their curvature radius -- this distinction will be irrelevant to our work.)

The main outcome of our work is that a thin wall propagating within an axion background experiences a $\gamma^2$-enhanced pressure if the axion decay constant changes across the interface. Parametrically, 
\be \label{eq:P_intro}
    \mathcal{P} \sim \gamma^2 \left( \frac{\Delta f^2}{f^2} \right)^2 \rho_\theta \times
    \begin{cases} m^2 / H^2 & \text{for} \quad m/H \ll 1 \\ 1 & \text{for} \quad m/H \gg 1 \end{cases} \ ,
\ee
where $\rho_\theta$ refers to the axion energy density at the relevant epoch, and $H$ the corresponding Hubble rate. During the frozen regime, this effect is suppressed by the ratio $m^2 / H^2 \ll 1$, and disappears in the massless limit. The backreaction on the axion field leads to a small component of axions becoming relativistic, with Lorentz factor $\gamma_\text{shell} \simeq 2 \gamma^2$. In some cases, these might remain relativistic until late times, contributing to $\Delta N_\text{eff}$ and potentially surviving as a dark radiation component at the present epoch. We emphasize that the effect of Eq.\eqref{eq:P_intro} is only present during the thin-wall regime, and so it is a transitory effect if the wall can continue to accelerate to $\gamma$-factors that exceed the thin-wall bound, $\gamma \sim (m L)^{-1}$. Interestingly, ALPs are often light or ultra-light, with tiny masses that break the axion shift-symmetry by (often exponentially) small non-perturbative effects, while the thickness of domain walls or phase-transition bubble walls is set by high scale physics entirely independent of the axion mass. As a result, it is often naturally the case that $m L \lll 1$ and so the `transitory' thin-wall regime can encompass a very large range of $\gamma$-factors, including far ultra-relativistic wall motion.

Finally, before we move on to outline the structure of this paper, it is worth clarifying two related and possibly confusing apparent differences between the bubble walls of a first order phase transition and the domain walls arising from spontaneous breaking of a discrete global symmetry. For the former, the absence of a global symmetry between the two sides of the interface leads to the expectation of $\Delta f^2 \neq 0$ being generic, e.g.~$\Delta f^2 / f^2 \simeq \kappa v^2 / f^2$ would follow  from Eq.\eqref{eq:dim6} if $\langle \phi \rangle$ changes from zero to $v$ across the wall.  For the latter, on the contrary, one would expect $\Delta f^2 = 0$ from Eq.\eqref{eq:dim6} if the discrete global symmetry is exact, e.g.~if $\langle \phi \rangle$ changes from $+v$ to $-v$ across the interface. In practice, though, a discrete global symmetry that is spontaneously broken in the early universe can only be approximate so as to avoid the so-called `domain wall problem' \cite{Zeldovich:1974uw}, and therefore non-zero $\Delta f^2$ is a general expectation in this case too. Moreover, discrete symmetry-breaking operators such as $\mu \phi (\partial_\mu \theta)^2$ may also be present, leading to a change in the axion decay constant across the wall $\Delta f^2 \simeq 2 \mu v$, even if we ignore other explicit symmetry-breaking effects. Second, the bubble walls formed during a phase transition expand fueled by the difference in energy densities between the vacua at either side of the wall and, as a result, they can easily reach relativistic velocities. A domain wall network, on the other hand, spends most of its life being non-relativistic. However, in the presence of global symmetry-breaking effects the network eventually collapses, and the walls can reach large $\gamma$-factors during this collapse phase. Even ignoring symmetry-breaking dynamics, there exists exceptions to the expectation that domain walls evolve non-relativistically. For example, walls formed with a spherical topology naturally approach relativistic velocities during their inward collapse driven by their surface tension, with interesting implications for the production of a population primordial black holes as well as higher frequency gravitational waves~\cite{Vachaspati:2017hjw,Ferrer:2018uiu,Gelmini:2022nim,Gouttenoire:2023gbn,Ferreira:2024eru,Dunsky:2024zdo}. The $\gamma^2$-enhanced friction described in this work can then also be relevant to these cases.

The rest of this paper is organized as follows. In section~\ref{sec:idea} we discuss the evolution of both planar and non-planar walls that interact with an axion background via a change in decay constant. We focus on the thin-wall regime and derive an expression for the pressure exerted on a relativistic wall, given in Eq.\eqref{eq:P2}, and clarify the role of the scale set by the wall thickness. We outline some of the potential phenomenological implications of this effect in section~\ref{sec:pheno}, including a modification of the dynamics of the bubble walls during a first order phase transition, and the potential presence of a relativistic axion component at the current epoch. Finally, section~\ref{sec:conclusions} briefly summarizes our conclusions, and appendices \ref{app:deltareg}-\ref{sec:Axion_Higgs_Portal_constraints} contain additional supplementary material.

Our work extends and generalizes some of the results presented in our previous paper~\cite{GarciaGarcia:2022yqb}, which focused on the effect of interactions between bubble walls and massive dark photons. More generally, understanding the dynamics of bubble walls and other extended defects during a cosmological phase transition is a longstanding problem that has received much attention in the past~\cite{Turok:1992jp,Dine:1992wr,Liu:1992tn,Arnold:1993wc,Moore:1995ua,Moore:1995si} and also more recently~\cite{Bodeker:2009qy,Espinosa:2010hh,Leitao:2015ola,Bodeker:2017cim,Azatov:2020ufh,Ai:2021kak,Mancha_2021,Bigazzi:2021ucw,Bea:2021zsu,Gouttenoire:2021kjv,Azatov:2022tii,Laurent:2022jrs,DeCurtis:2022hlx,Wang:2022txy}.

\section{Axion-Wall Evolution}
\label{sec:idea}

In \ref{sec:planar} we discuss the evolution of a planar wall interacting with a classical axion background, focusing on the regime where the thickness of the wall can be neglected. The main result of this subsection is our derivation of the pressure experienced by a moving wall, and it is given in Eq.\eqref{eq:P2}.
In \ref{sec:bubble} we go beyond the planar wall approximation, and discuss the effect of a finite curvature radius on the axion-wall system.
In \ref{sec:qm} we present an alternative derivation of the pressure experienced by a planar wall that clarifies the role of the scale set by the wall thickness, and that can be extended to scenarios where the change in axion decay constant need not be small.
Finally, in \ref{sec:setup} we comment on the regime of validity of the EFT describing the axion-wall system.
Throughout this section, we neglect the non-trivial evolution of the background spacetime, and consider the evolution of the axion-wall system in an otherwise Minkowski background.
This will turn out to be a good approximation to describe the dynamics of the system over time scales $\Delta t \ll H^{-1}$, which will be the context of section~\ref{sec:pheno}.

\subsection{Planar thin wall on an axion background}
\label{sec:planar}

The effective action describing the dynamics of a rigid, planar wall following a trajectory $z = z_0(t)$ is given by
\begin{equation} \label{eq:S_wall}
    S_\text{wall}= A \int dt \left\{  - \sigma \sqrt{1 - \dot z_0(t)^2} + z_0(t) \Delta V \right\} \ ,
\end{equation}
where $A$ is the wall area and $\sigma$ its surface tension.
The first term above corresponds to the Nambu-Goto action of a rigid thin wall, and the second term captures the possibility of a difference in energy densities, $\Delta V$, between the vacua at either side of the interface. The equation of motion (EOM) for the wall following from Eq.\eqref{eq:S_wall} reads
\be \label{eq:EOM_wall}
	\ddot z_0 (t) = \frac{\Delta V}{\gamma(t)^3 \sigma}
    \qquad \text{with} \qquad \gamma(t) \equiv \left( 1 - \dot z_0(t)^2 \right)^{-1/2} \ .
\ee

Interactions between the wall and other dof will modify Eq.\eqref{eq:EOM_wall}. Here, we will focus on the effect of an axion background that interacts with the wall via a change in decay constant. The corresponding effective action for the axion field is given by
\be \label{eq:S_theta}
    S_\text{axion} = \frac{1}{2} \int d^4 x \left\{ f^2_\text{w} (x) (\partial_\mu \theta)^2 - \Lambda^4 \theta^2 + \dots \right\} \ ,
\ee
where $\Lambda^4$ captures the breaking of the axion shift-symmetry and the dots denote interactions other than quadratic.
The classical EOM describing the evolution of the axion field following from Eq.\eqref{eq:S_theta} reads
\be \label{eq:EOM_theta}
    \partial_\mu( f_\w^2 (x) \partial^\mu \theta) + \Lambda^4 \theta = 0 \ .
\ee
In the thin-wall limit, we will approximate the effective axion decay constant as
\be \label{eq:f2w}
    f^2_\text{w} (t,z) = f^2 + \Delta f^2 {\rm H} (\gamma(t) (z_0(t) - z)) \ ,
\ee
where ${\rm H}$ is the Heaviside step function centered at the location of the wall. In keeping with our discussion in and around Fig.~\ref{fig:setup}, we consider a wall moving to the right ($\dot z_0 (t) > 0$), and such that $f_\w^2(x)$ takes values $f^2$ and ${\tilde f}^2 = f^2 + \Delta f^2$ to the right ($z>z_0(t)$) and to the left ($z<z_0(t)$) of the wall respectively.

The axion-wall interaction of Eq.\eqref{eq:S_theta} further modifies the wall EOM, and Eq.\eqref{eq:EOM_wall} gets replaced by
\be \label{eq:EOM_wall_P}
	\ddot z_0(t) = \frac{\Delta V - \mathcal{P} (t)}{\gamma(t)^3 \sigma} \ ,
\ee
with $\mathcal{P}(t)$ given by
\be \label{eq:P_wall}
	\mathcal{P} (t) = - \frac{1}{A} \frac{\delta S_\text{axion}}{\delta z_0 (t)} = - \frac{\Delta f^2}{2} (\partial_\mu \theta)^2 \Big|_{z=z_0(t)} \ .
\ee
Eq.\eqref{eq:EOM_theta} and \eqref{eq:EOM_wall_P}-\eqref{eq:P_wall} describe the joint evolution of the axion-wall system.

In the following, we will proceed to solve the axion-wall equations of motion perturbatively, expanding $\theta$ as the sum of a homogeneous background, $\bar \theta(t)$, and a perturbation created by the presence of the wall, i.e.~
\be
    \theta (t,z) = \bar \theta (t) + \delta \theta(t,z) \ ,
\ee
with initial conditions at the moment the wall nucleates, $t=t_n$, given by $\delta \theta (t_n, z) \equiv 0$.
The right-hand-side of Eq.\eqref{eq:P_wall} can then be conveniently divided in two different pieces, as follows:
\begin{align}
	\mathcal{P} (t) & =
    - \frac{\Delta f^2}{2} {\dot{\bar \theta}}^{\, 2}
    - \Delta f^2 \left. \left( \dot{\bar \theta} \, \dot{\delta \theta} + \frac{1}{2} \left( \partial_\mu \delta \theta \right)^2 \right) \right|_{z=z_0(t)} \ . \label{eq:P_wall_3}
\end{align}
The first term in Eq.\eqref{eq:P_wall_3} corresponds to the pressure exerted on the wall due to an existing axion background, and is independent of the $\gamma$-factor of the expanding wall. On the other hand, in the regime $\gamma \gg 1$, both terms in parenthesis are of $\mathcal{O} (\gamma^2)$ and therefore easily provide the dominant contribution to $\mathcal{P} (t)$ in the relativistic limit. At leading order in the presence of the wall, i.e.~in the small parameter $\Delta f^2 / f^2 \ll 1$, the term $\propto \dot{\bar \theta} \, \dot{\delta \theta}$ is dominant.
In total, including terms up to second order in the ratio $\Delta f^2 / f^2$, Eq.\eqref{eq:P_wall_3} is well-approximated by
\begin{align}
	\mathcal{P} (t) & \simeq - \frac{\Delta f^2}{2} {\dot{\bar \theta}}^{\, 2}
    \underbrace{ - \left. \Delta f^2 \dot{\bar \theta} \, \dot{\delta \theta} \right|_{z=z_0(t)} }_{ \text{thin wall}} \ . \label{eq:P_wall_approx} 
\end{align}
There is an important additional distinction between these two terms. Whereas the first term is present whenever $\gamma \gg 1$, the second term is only present in the thin-wall regime. As it will become clear in \ref{sec:qm}, the latter requires that the $\gamma$-factor of the expanding wall does not parametrically exceed the combination $(m L)^{-1}$, where $m$ refers to the axion mass and $L$ the wall thickness (in the wall rest frame). For values of $\gamma$ such that the axion de Broglie wavelength no longer exceeds the wall thickness, i.e.~$\gamma \gtrsim (mL)^{-1}$, the thin-wall approximation no longer holds: the second term in Eq.\eqref{eq:P_wall_approx} `turns off' and the term $\propto \Delta f^2 {\dot{\bar \theta}}^{\, 2}$ is left as the leading contribution to the wall pressure. We will continue to focus on the thin-wall regime during the rest of this subsection, in which case the second term in Eq.\eqref{eq:P_wall_approx} provides a good approximation to the overall pressure in the relativistic limit.

We will now proceed to solve the EOM for $\delta \theta (t,z)$ that follows from Eq.\eqref{eq:EOM_theta} perturbatively in the presence of the wall.
At $\mathcal{O} \left( \Delta f^2 / f^2 \right)$, we have:
\be \label{eq:EOM_theta1}
    (\Box + m^2) \delta \theta^{(1)} = j^{(1)} \ ,
\ee
with
\begin{align}
    j^{(1)} & = - \frac{\dot f^2_\text{w}}{f^2} \dot {\bar \theta} + m^2 \frac{f_\w^2 - f^2}{f^2} \bar \theta \label{eq:j1_full} \\
            & \simeq - \frac{\dot f^2_\text{w}}{f^2} \dot {\bar \theta} \qquad \textrm{($\gamma \gg 1$)} \ . \label{eq:j1}
\end{align}
For a thin wall (recall Eq.\eqref{eq:f2w}) the first term in Eq.\eqref{eq:j1_full} is a delta-function source localized on the wall, whereas the second term is proportional to the Heaviside step-function of Eq.\eqref{eq:f2w}. In the regime $\gamma \gg 1$, it is the first term that dominates, as indicated in Eq.\eqref{eq:j1}. The solution to Eq.\eqref{eq:EOM_theta1} can then be written in closed form as follows
\be \label{eq:theta1_sol}
    \delta \theta^{(1)} (t,z) = \iint dt' dz' G_\oned (t-t',z-z')j^{(1)} (t',z') \ ,
\ee
where $G_\oned$ is the one-dimensional retarded Green's function for the Klein-Gordon operator, given by the standard expression
\be
    G_\oned (t,z) = \frac{1}{2} {\rm H}(t) {\rm H}(t^2 - z^2) J_0 \left( m \sqrt{t^2 - z^2} \right) \ ,
\ee
with $J_0$ a Bessel function of the first kind. Evaluating Eq.\eqref{eq:theta1_sol} with $j^{(1)}$ as given in Eq.\eqref{eq:j1} one finds
\be \label{eq:theta1_wall}
    \delta \theta^{(1)} (t,z)  \simeq - \frac{\Delta f^2}{2 f^2} \int_{t_n}^t dt' \dot z_0(t') \dot {\bar \theta} (t') {\rm H}(s^2) J_0 (m s) \Big|_{z'=z_0(t')} \ ,
\ee
where $s \equiv \sqrt{(t-t')^2 - (z-z')^2}$. In turn, in the regime $\gamma \gg 1$:
\begin{align} \label{eq:dottheta1_wall}
    \delta \dot \theta^{(1)} \Big|_{z=z_0(t)} \simeq - \frac{\Delta f^2}{2 f^2} \gamma^2 \dot {\bar \theta}^{\, 2} \ .
\end{align}
In total, the leading contribution to the pressure on the expanding wall can be written as (recall Eq.\eqref{eq:P_wall_approx})
\be \label{eq:P2}
   \boxed{ \mathcal{P} \simeq - \frac{\Delta f^2}{2} \dot {\bar \theta}^{\, 2}
   + \underbrace{ \frac{\gamma^2}{2} \left( \frac{\Delta f^2}{f^2} \right)^2 f^2 \dot {\bar \theta}^{\, 2} }_{\text{thin wall}} } \ ,
\ee
where, as mentioned earlier, the second term above is present provided the thickness of the wall can be neglected.
Eq.\eqref{eq:P2} is the main result of this section, and we emphasize it is valid for an arbitrary wall trajectory $z_0(t)$. Despite the additional $\Delta f^2 / f^2$ suppression, the second term will dominate provided $\gamma \gg ( \Delta f^2 / f^2 )^{-1/2}$, and therefore easily provides the dominant contribution to the wall pressure in the relativistic limit.

As a toy example, consider an axion background of the form $\bar \theta (t) = \mathcal{C} \cos (m t)$, with $\mathcal{C} = \text{constant.}$
Averaging over timescales large compared to $m^{-1}$, Eq.\eqref{eq:P2} can be written as
\be \label{eq:Pex_average}
    \langle \mathcal{P} \rangle \simeq - \frac{\Delta f^2}{2 f^2} \rho_\theta + \frac{\gamma^2}{2} \left( \frac{\Delta f^2}{f^2} \right)^2 \rho_\theta \ ,
\ee
where $\rho_\theta \equiv \frac{1}{2} \mathcal{C}^2 m^2 f^2$ is the axion energy density. As we will discuss in \ref{sec:gammaeq}, Eq.\eqref{eq:Pex_average} provides a good description of the pressure experienced by a relativistic thin wall in the presence of an oscillating axion background in the early universe.

A final comment before we move on. Although we have derived our results in a basis where the axion field is dimensionless, one can equivalently work in a basis where the axion field is canonically normalized, i.e.~$a \equiv f_\text{w} (x) \theta$. In this case, Eq.\eqref{eq:S_theta} and Eq.\eqref{eq:EOM_theta} read
\be
    S_\text{axion} = \frac{1}{2} \int d^4 x \left\{ (\partial_\mu a)^2 - m^2_\text{eff}(x) a^2 + \dots \right\} \ ,
\ee
and
\be
    (\Box + m^2_\text{eff}(x)) a = 0 \ ,
\ee
respectively, and the `effective' axion mass $m^2_\text{eff}(x)$ is given by
\be \label{eq:m2eff}
    m^2_\text{eff} (x) = \frac{\Lambda^4}{f_\text{w}^2} - \frac{1}{2} \frac{\Box f_\text{w}^2}{f_\text{w}^2} + \frac{1}{4} \left( \frac{\partial_\mu f_\text{w}^2}{f_\text{w}^2} \right)^2 \ .
\ee
The first term above represents a step-function change on the axion mass -- following from its non-trivial dependence on the decay constant (remember Fig.~\ref{fig:setup} and surrounding discussion). This term is responsible for the first term in Eq.\eqref{eq:Pex_average}, which more generally can be written as $\frac{\Delta m^2}{2 m^2} \rho_\theta$, and $\Delta m^2 / m^2 \simeq - \Delta f^2 / f^2$ in our particular example. The last two terms in Eq.\eqref{eq:m2eff}, on the other hand, are highly localized at the location of the wall, and are responsible for the $\gamma^2$-enhanced contribution to the wall pressure.
If only the axion mass -- and not  the decay constant -- varied across the interface then this $\gamma^2$-enhanced contribution would absent.

\subsection{Beyond planar walls}
\label{sec:bubble}

So far, we have focused on the limit where the expanding domain or bubble walls can be treated as planar. Whereas this may be appropriate to describe the evolution of large domain walls, it is less clear that it correctly captures the behavior of bubble walls whose initial curvature radius is $\ll H^{-1}$. We now address this question.

Extending the analysis of section~\ref{sec:planar} to the case of a thin, spherical bubble wall is relatively straightforward. In the absence of interactions, the effective action for a spherical interface reads (cf.~Eq.\eqref{eq:S_wall})
\be \label{eq:S_bubble}
	S_\text{wall} = \int dt \left\{ - 4 \pi \sigma R(t)^2 \sqrt{1 - \dot R(t)^2} + \frac{4 \pi}{3} R(t)^3 \Delta V \right\} \ ,
\ee
and the corresponding EOM is given by
\be \label{eq:EOM_bubble}
	\ddot R(t) + \frac{2}{R(t) \gamma(t)^2} = \frac{\Delta V}{\sigma \gamma(t)^3} \ ,
\ee
with $\gamma(t) = ( 1 - \dot R(t)^2 )^{-1/2}$. As it is well-known, the solution to Eq.\eqref{eq:EOM_bubble} with vanishing initial velocity is given by $R(t) = ( \Delta t^2 +R_n^2 )^{1/2}$, with $\Delta t \equiv t-t_n$ the time elapsed after the appearance of the bubble and $R_n = 3 \sigma / \Delta V$ its initial radius. In vacuum, the bubble wall then follows a trajectory of constant proper acceleration, with the corresponding $\gamma$-factor $\gamma(t) = R(t) / R_n$ ever increasing as the bubble grows~\cite{Coleman:1977py}.

As before, we will now consider interactions of the form Eq.\eqref{eq:S_theta}, except the effective axion decay constant now takes the form
\begin{align} \label{eq:f2_bubble}
	f^2_\w (t, r) = f^2 + \Delta f^2 \, {\rm H} ( \gamma (t) (R(t) - r) ) \ ,
\end{align}
so that $f_\w^2(x)$ takes values $f^2$ and ${\tilde f}^2 = f^2 + \Delta f^2$ outside ($r>R(t)$) and inside ($r<R(t)$) of the vacuum bubble respectively. The wall EOM now reads
\be \label{eq:dSdR}
	\ddot R(t) + \frac{2}{R(t) \gamma(t)^2} = \frac{\Delta V - \mathcal{P}(t)}{\sigma \gamma(t)^3} \ ,
\ee
with
\be \label{eq:P_bubble}
	\mathcal{P} (t) = - \frac{1}{4 \pi R(t)^2} \frac{\delta S_\text{axion}}{\delta R (t)} = - \frac{\Delta f^2}{2} (\partial_\mu \theta)^2 \Big|_{r=R(t)} \ .
\ee
So far, this closely mirrors the results obtained in~\ref{sec:planar} for a planar wall.

To evaluate Eq.\eqref{eq:P_bubble} in perturbation theory we proceed as in section~\ref{sec:planar}, except now
\be \label{eq:theta1_bubble}
    \delta \theta^{(1)} (t,r) = \iint dt' d^3 {\bf r'} \, G_\text{3d} (t-t', |{\bf r} - {\bf r'}|) j^{(1)} (t', {\bf r'}) \ ,
\ee
where $ G_\text{3d}$ is the three-dimensional retarded Green's function, i.e.~
\be
    G_\text{3d} (t, r) = {\rm H}(t) \left\{ \frac{\delta(t^2 - r^2)}{2\pi} - \frac{m^2}{4\pi} {\rm H}(t^2 - r^2) \frac{J_1 ( m\sqrt{t^2-r^2} )}{m\sqrt{t^2-r^2}} \right\} \ ,
\ee
with $J_1$ a Bessel function of the first kind, and the source $j^{(1)}$ as in Eq.\eqref{eq:j1_full}-\eqref{eq:j1}, except with $f^2_\text{w}$ as given in Eq.\eqref{eq:f2_bubble}. As before, neglecting the last term in Eq.\eqref{eq:j1_full} and performing the integral over ${\bf r'}$, we find
\be \label{eq:delta1_bubble}
    \delta \theta^{(1)} (t,r) \simeq    - \frac{\Delta f^2}{2 f^2} \int_{t_n}^t dt' \frac{r'}{r} \dot R(t') \dot {\bar \theta} (t') {\rm H}(s^2) \left( J_0 (m s) - {\rm H}({\tilde s}^2) J_0 (m \tilde s) \right) \Big|_{r' = R(t')} \ ,
\ee
where $s \equiv \sqrt{(t-t')^2 - (r-r')^2}$ and $\tilde s \equiv \sqrt{(t-t')^2 - (r+r')^2}$.
Eq.\eqref{eq:delta1_bubble} differs from Eq.\eqref{eq:theta1_wall} in a number of ways. Most notably, the term in Eq.\eqref{eq:delta1_bubble} contains an additional piece $\propto {\rm H}({\tilde s}^2)J_0 (m \tilde s)$ that is absent in Eq.\eqref{eq:theta1_wall}. This term captures contributions to $\delta \theta^{(1)} (t,r)$ that are sourced at time $t'$ from the furthest point on the bubble wall (a distance $r' + r$ from the point at which we are evaluating the perturbation), and it interferes destructively with the term $\propto J_0 (ms)$. The geometric factor $r' / r$ in Eq.\eqref{eq:delta1_bubble} further differentiates the planar vs spherical wall calculations.

Interestingly, the relevant quantity that enters into Eq.\eqref{eq:P_bubble} in the limit $\gamma \gg 1$ remains unaffected by the curvature of the wall. At the location of the wall, $r=R(t)$, one has $s^2 > 0$ and $\tilde s^2 < 0 $, so there is no contribution to either $  \delta \theta^{(1)}$ or $  \dot{\delta \theta}^{(1)}$ from the second term in Eq.\eqref{eq:delta1_bubble}. Proceeding much like in section~\ref{sec:planar}, one finds
\be
    \dot{\delta \theta}^{(1)} \Big|_{r = R(t)} \simeq - \frac{\Delta f^2}{2 f^2} \gamma^2 \dot {\bar \theta} \ .
\ee
In turn, including terms up to second order in $\Delta f^2 / f^2$, Eq.\eqref{eq:P_bubble} reads
\be
    \mathcal{P} \simeq - \frac{\Delta f^2}{2} {\dot {\bar \theta}}^{\, 2} + \frac{\gamma^2}{2} \left( \frac{\Delta f^2}{f^2} \right)^2 f^2 \dot {\bar \theta}^2 \ ,
\ee
in complete analogy with Eq.\eqref{eq:dottheta1_wall}-\eqref{eq:P2}. Thus, the pressure experienced by the expanding bubble wall in the ultra-relativistic limit remains well-approximated by Eq.\eqref{eq:P2}.

We emphasize that this conclusion is not necessarily true more generally. For instance, consider a nearly frozen axion field such that $\dot {\bar \theta} \simeq \mathcal{C} m^2 t$, with $m t \ll 1$ and $\mathcal{C} = \text{constant.}$  At the location of the interface, $r=R(t)$, the size of the perturbation induced by an accelerated bubble wall is given by
\begin{align}
    \delta \theta^{(1)} \Big|_{r=R(t)}
    &\simeq    - \frac{\Delta f^2}{2 f^2} \int_{t_n}^t dt' \frac{r'}{r} \dot R(t') \dot {\bar \theta} (t') J_0 (m s) \Big|_{r' = R(t') , \, r = R(t)} \\
    & \simeq - \frac{\Delta f^2}{2 f^2} \frac{\mathcal{C} m^2}{R(t)} \int_{t_n}^t dt' (t' - t_n) t' \label{eq:dtheta1_frozen} \\
    & \simeq - \frac{\mathcal{C}}{4} \frac{\Delta f^2}{f^2} m^2 t ( t - t_n) \label{eq:dtheta1_frozen_approx}
    \ ,
\end{align}
where in Eq.\eqref{eq:dtheta1_frozen_approx} we have taken $t - t_n \ll t$ for illustration, and $R(t) \simeq t -t_n \gg R_n$. In contrast, the corresponding perturbation generated by a planar wall following the same accelerated trajectory is given by
\begin{align}
    \delta \theta^{(1)} \Big|_{z=z_0(t)}
    &\simeq    - \frac{\Delta f^2}{2 f^2} \int_{t_n}^t dt' \dot z_0(t') \dot {\bar \theta} (t') J_0 (m s) \Big|_{z' = z_0(t') , \, z = z_0(t)} \\
    & \simeq - \frac{\Delta f^2}{2 f^2} \mathcal{C} m^2 \int_{t_n}^t dt' t'  \\
    & \simeq - \frac{\mathcal{C}}{2} \frac{\Delta f^2}{f^2} m^2 t ( t - t_n) 
    \ ,
\end{align}
which differs from Eq.\eqref{eq:dtheta1_frozen_approx} by a factor of 2 due to the different geometry of the two configurations.

\subsection{Particle interpretation and wall thickness}
\label{sec:qm}

In the regime where axion-wall interactions can be thought of as the result of scattering between individual particles and the wall, an alternative, simpler, derivation of the pressure experienced by the interface is possible. This will be appropriate, for example, to describe the effect of an oscillating axion field, and we will show the complete equivalence between the two pictures.  On the other hand, the frozen axion regime requires the full field-theoretic treatment developed in section~\ref{sec:planar}. Despite this, the particle interpretation will illuminate the origin of the $\gamma^2$-enhancement in the wall pressure, and it will help clarify the crucial role of the scale set by the wall thickness.

Consider a single axion particle impinging on the wall. The probability for it to transmit or reflect can be obtained by solving its EOM Eq.\eqref{eq:EOM_theta}, which for a thin wall reads
\be \label{eq:EOM_theta_thin}
    (\Box + \tilde m^2) \theta = 0 \quad \text{for} \quad z < z_0(t)
    \qquad \text{and} \qquad
     (\Box + m^2) \theta = 0 \quad \text{for} \quad z > z_0(t) \ ,
\ee
with $\tilde m^2 = \Lambda^4 / \tilde f^2$ and $m^2 = \Lambda^4 / f^2$, as discussed in and around Fig.~\ref{fig:setup}. Assuming that the wall propagates with constant velocity, i.e.~$\dot z_0 \equiv v = \text{cst.}$, the solution to Eq.\eqref{eq:EOM_theta_thin} with the appropriate scattering boundary conditions can be conveniently written in terms of plane waves, as follows
\be \label{eq:theta_sol}
	\theta(\hat t, \hat z) = \mathcal{N} \, e^{- i \omega \Delta \hat t }
	\begin{cases} 	e^{-i k \hat z} + r_\theta e^{i k \hat z} & \quad \text{for} \ \ \quad \hat z > 0 \\
				t_\theta e^{-i \tilde k \hat z} & \quad \text{for} \ \ \quad \hat z < 0
	\end{cases} \ ,
\ee
where $\Delta \hat t \equiv \gamma (\Delta t - vz) $ and $\hat z \equiv \gamma (z - v \Delta t)$ are coordinates appropriate to the rest frame of the wall (located at $\hat z = 0$), $\mathcal{N}$ is an unimportant normalisation factor, and it is implicit that only the real part is taken on the right-hand-side of Eq.\eqref{eq:theta_sol}. $\omega = \gamma m$ is the energy of the incoming axion particle in the rest frame of the wall, $k = \sqrt{\omega^2 - m^2} = v \gamma m$ and $\tilde k = \sqrt{\omega^2 - {\tilde m}^2}$.

To obtain the reflection and transmission coefficients, $r_\theta$ and $t_\theta$, Eq.\eqref{eq:EOM_theta_thin} must be supplemented with the appropriate matching conditions for $\theta$ and its first derivative at the location of the interface. Integrating Eq.\eqref{eq:EOM_theta} twice across the location of the wall we find that these are given by
\be
\label{eq:FchangeBCs}
	\theta \Big|_{\hat z = 0^-}^{\hat z = 0^+} = 0
	\qquad \text{and} \qquad
	f_\w^2 (\hat z) \partial_{\hat z} \theta \Big|_{\hat z = 0^-}^{\hat z = 0^+} = 0 \ .
\ee
In particular, $\partial_{\hat z} \theta$ features a step-discontinuity at $\hat z = 0$. In turn:
\be \label{eq:RandT_f2}
    r_\theta = \frac{f^2 k - {\tilde f}^2 \tilde k}{f^2 k + {\tilde f}^2 \tilde k}
    \qquad \text{and} \qquad
	t_\theta = 1 + r_\theta = \frac{2 f^2 k}{f^2 k + {\tilde f}^2 \tilde k} \ .
\ee
In the relativistic regime, $\omega \gg m, \tilde m$, the reflection probability is therefore given by
\be \label{eq:R}
    \mathcal{R} \equiv |r_\theta|^2 \simeq \left( \frac{f^2 - \tilde f^2}{f^2 + \tilde f^2} \right)^2 \simeq \left( \frac{\Delta f^2}{2 f^2} \right)^2 \ ,
\ee
where in the last step we have further assumed that $\Delta f^2 / f^2 \ll 1$. The reflection probability stays constant in the relativistic limit as a direct result of the discontinuity of $\partial_{\hat z} \theta$ at $\hat z = 0$.

The momentum transferred to the wall from transmitted and reflected axions exerts a force on the moving interface. Per unit area of the wall, this is given by the average axion flux hitting the wall times the corresponding momentum transfer from reflected and transmitted particles, i.e.~
\begin{align}
\label{eq:ParticlePressure}
	\frac{F}{A} = \underbrace{ v \gamma n_\theta }_\text{flux} \times \underbrace{ \Big( \mathcal{R} \, 2 k + \mathcal{T} \, (k - \tilde k) \Big) }_\text{momentum transfer} \ ,
\end{align}
where $n_\theta$ is the number density of axions (in their own rest frame), and $\mathcal{R} = |r_\theta|^2$ and $\mathcal{T} = \frac{\tilde k}{k} |t_\theta|^2$ are reflection and transmission probabilities satisfying $\mathcal{R} + \mathcal{T} = 1$. In the relativistic limit, and to leading order in the ratio $\Delta f^2 / f^2$ for each non-negative power of $\gamma$, we find
\be \label{eq:F/A}
	\frac{F}{A} \simeq - \frac{\Delta f^2}{2 f^2} \rho_\theta + \underbrace{ \frac{\gamma^2}{2} \rho_\theta \left( \frac{\Delta f^2}{f^2} \right)^2 }_{\text{thin wall}} \ ,
\ee
with $\rho_\theta \equiv m n_\theta$ the axion energy density. The first and second terms above originate in the momentum transferred by transmitted and reflected axions respectively. As anticipated, this agrees with our previous result in Eq.\eqref{eq:Pex_average}.

We can actually show that the two methods match at all orders. For a wall travelling at constant speed we can take (the real part of) Eq.\eqref{eq:theta_sol} to describe the steady-state real solution to its EOM, rather than a single particle wavefunction. $\mathcal{N}$ then defines the amplitude of oscillation and fixes the axion number density. Having an exact solution, we can then compute pressure exactly in terms of its field theoretic expression Eq.\eqref{eq:P_wall} evaluated at the position of the wall. However, the latter is actually ill-defined due to the discontinuity of the solution's first derivative in $z$, made explicit in Eq.\eqref{eq:FchangeBCs}. Evaluating the field theoretic expression fully requires the non-trivial task of regulating the step function, as we do explicitly in appendix~\ref{app:deltareg}, showing it matches Eq.\eqref{eq:ParticlePressure}. In this instance, the particle derivation is thus much simpler. 

\paragraph{Beyond thin walls:}
Let us move beyond the thin-wall approximation, and consider a wall profile that generalizes that of Eq.\eqref{eq:f2w} such that the axion decay constant varies smoothly across the interface, that is
\be
    f^2_\text{w} (\hat z) = f^2 + \Delta f^2 {\rm H} (\hat z) \quad \xrightarrow{\text{finite thickness}} \quad f^2 + \Delta f^2 \text{H}_L (\hat z) \ ,
\ee
with e.g.~
\be
    \text{H}_L (\hat z) = \frac{1}{2} \left( 1 + \tanh(\hat z / L) \right) \ .
\ee
When $\Delta f^2 / f^2 \ll 1$, the reflection coefficient is always tiny, and so we can obtain an analytic expression for the reflection probability in the Born approximation. Proceeding as in \cite{GarciaGarcia:2022yqb} (see section~III.C), we find
\begin{align}
    \mathcal{R}_\text{Born}
        & \simeq \left( \frac{\Delta f^2}{2 f^2} \right)^2 \left| \int_{-\infty}^{+\infty} d \hat z \, e^{2 i k \hat z} \, \text{H}'_L (\hat z) \right|^2 \\
        & \simeq \left( \frac{\Delta f^2}{2 f^2} \right)^2 \frac{\pi^2 (k L)^2}{\sinh (\pi k L)^2} \ .
\end{align}
This expression reproduces Eq.\eqref{eq:R} in the regime $k L \ll 1$, whereas $\mathcal{R}_\text{Born} \propto e^{-2 \pi k L}$ at energies $k \simeq \omega \simeq \gamma m \gg L$. In this `thick wall' regime the reflection probability decays exponentially as $\gamma$ increases, and so effectively the $\gamma^2$-enhanced contribution to the wall pressure sharply turns off for $\gamma$-factors $\gtrsim (m L)^{-1}$.

\subsection{Regime of validity of the effective field theory}
\label{sec:setup}

So far we have focused on interactions between a two-dimensional wall-like defect and a classical axion background without considering the potential limitations of this effective treatment. We now comment on this question.

In general, the behavior of the scalar order parameter describing spontaneous symmetry breaking is governed by an effective Lagrangian of the form
\be \label{eq:Lphi}
    \mathcal{L_\phi} = \frac{1}{2} (\partial_\mu \phi)^2 - V_T (\phi) + \dots \ ,
\ee
with $V_T (\phi)$ an effective potential that in general receives corrections at both zero and finite temperature. Here, we have been concerned with situations where $V_T (\phi)$ features a discrete number of metastable and stable vacua, and there exist field configurations that interpolate between them, $\langle \phi \rangle= v(x)$, corresponding to two-dimensional wall-like defects.  For our purposes, this could correspond to a spherical bubble wall nucleated during a first order phase transition, or a nearly planar domain wall arising from the spontaneous breaking of a discrete global symmetry.  The physical properties of the wall, such as its tension and thickness are primarily set by the parameters appearing in the effective potential of Eq.\eqref{eq:Lphi}. As anticipated in section~\ref{sec:intro}, ALPs will generally feature shift-symmetry preserving couplings with the scalar order parameter, as illustrated in Eq.\eqref{eq:dim6}. These couplings have minimal effect on the static properties of the interface, but play a vital role in the later evolution of the axion-wall system.

Obviously, Eq.\eqref{eq:dim6} is an effective interaction and implies an upper bound on the cutoff scale of the EFT, of the form
\be \label{eq:cutoff}
    \Lambda \lesssim \frac{4 \pi f}{\sqrt{\kappa}}
            = \frac{4 \pi v}{\sqrt{\kappa v^2 / f^2}} \ .
\ee
If $\kappa v^2 / f^2 \ll 1$, this cutoff scale can be well above the typical scale of the $\phi$ sector, $v$, that forms part of the axion EFT.  This is the parametric regime that we will focus on in this work. In the case of bubble walls formed during a phase transition the quantity inside the square root is precisely the fractional change in the axion decay constant, i.e.~$\Delta f^2 / f^2 \simeq \kappa v^2 / f^2$. On the other hand, for domain walls formed during the spontaneous breaking of a discrete global symmetry, one may have $\Delta f^2 / f^2 \ll \kappa v^2 / f^2$, depending on the size of the \emph{explicit} breaking that leads to $\Delta f^2 \neq 0$ across the wall (recall discussion in section~\ref{sec:intro}).

To emphasize how generic interactions of the form Eq.\eqref{eq:dim6} are notice that a simple UV-completion is given by the following quartic interaction
\be \label{eq:dim4}
    \mathcal{L} \supset \frac{\eta}{4} \phi^2 |\Phi|^2 \ .
\ee
$\Phi$ refers to the complex scalar field, with vev $\langle |\Phi| \rangle = f / \sqrt{2}$, the phase of which is the axion. Integrating out the accompanying radial excitation at tree-level, one recovers Eq.\eqref{eq:dim6} with $\kappa \sim \eta f^2 / m_\rho^2$, and where $m_\rho$ is the mass of the radial mode. Eq.\eqref{eq:dim4} cannot be forbidden on the basis of symmetry, and therefore the natural expectation is for $\eta$ to be non-zero. As a result, shift-symmetry preserving interactions such as Eq.\eqref{eq:dim6} must be considered generic in any axion EFT.
\footnote{Other interactions are also generated after integrating out the radial mode of the UV theory and should therefore be expected to form part of the axion EFT. For example, axion self-interactions of the form $\mathcal{L}_\text{EFT} \supset \lambda_\theta (\partial_\mu \theta)^2 (\partial_\nu \theta)^2$ are generated with $\lambda_\theta \sim f^2 / m_\rho^2$ in this particular UV-completion. These can be relevant to the later evolution of the axion-wall system, as we will comment on~\ref{sec:tmunu}.}
We note that a hierarchy of scales $v \ll f$ would require a fine-tuning of the scalar potential for the order parameter, of size $\epsilon \equiv \frac{\lambda_\phi v^2}{\eta f^2} \sim \kappa^{-1} \frac{m_\phi^2}{m_\rho^2}$, where $\lambda_\phi$ refers to the typical quartic coupling in the $\phi$ sector, and $m_\phi^2 \sim \lambda_\phi v^2$ is the mass-squared of the scalar excitation accompanying the order parameter. Within this UV-completion, a natural hierarchy (i.e.~$\epsilon \gtrsim 1$) would thus require $\eta \lesssim m_\phi^2 / f^2$ or, equivalently, $\kappa \lesssim m_\phi^2 / m_\rho^2$. Other UV completions consistent with naturally large $\kappa$ are of course possible.

\section{Aspects of Phenomenology}
\label{sec:pheno}

In this section, we discuss some of the potential implications of the axion-wall interactions described in section~\ref{sec:idea} for the evolution of cosmological wall-like defects and a background axion field. In the early universe, the bubble walls of a first order phase transition are nucleated at rest, but can quickly reach relativistic velocities driven by the difference in energy densities $\Delta V$ between false and true vacua. To complete successfully and avoid a second inflationary period, a cosmological phase transition must complete within a fraction of the relevant Hubble time. Similarly, although a network of domain walls can evolve non-relativistically for long periods, the collapse phase during which the walls reach relativistic velocities must again be short-lived. Overall, the timescales over which we expect these wall-like defects to be relativistic in the early universe is at most of order $H^{-1}$, and typically $\Delta t \ll H^{-1}$. On these timescales, we therefore expect that neglecting the evolution of the background spacetime on the dynamics of the axion-wall system should be a decent approximation. This allows us to use the results derived in section~\ref{sec:idea}, with the classical axion background appropriate to the relevant epoch.

We will concentrate on the case where the mass of the axion field is independent of temperature. This is common in ALP extensions of the SM -- but crucially not true for the QCD axion, which is very sensitive to temperature at scales above $T_c \approx 150 \MeV$ \cite{GrillidiCortona:2015jxo,Borsanyi:2016ksw}. It is straightforward to generalise our results to this case.

For concreteness, we will focus on a cosmological axion background produced via the misalignment mechanism~\cite{Preskill:1982cy,Abbott:1982af,Dine:1982ah}. This follows whenever the PQ symmetry is broken during inflation, nor is it restored thereafter, i.e. $f \gg \max \left\{ H_{\rm I},T_{\rm rh} \right\}$, with $H_{\rm I}$ and $T_{\rm rh}$ the Hubble scale during inflation and the reheat temperature respectively. As long as $H_I\gg m$, at the end of inflation the axion field starts with a nearly homogeneous value $\theta_i$ (to be taken as a free, incalculable initial condition) and evolves in time according to the EOM:
\be \label{eq:scalarH}
    \ddot {\bar\theta} + 3 H \dot {\bar\theta} + V'(\bar\theta) = 0 \ .
\ee
Initially, the axion is frozen to its initial value $\bar\theta (t) \simeq \theta_i$ and contributes to the total energy density as an approximate form of dark energy. Eventually, as $H$ drops, the axion field begins oscillating and gravitates like pressureless matter. During a radiation- or matter-dominated era, and assuming $V(\theta_i) \simeq \frac{1}{2} f^2 m^2 \theta_i^2$ is a good approximation, the onset of oscillations occurs at a time $t \sim m^{-1}$, with $m$ referring to the axion mass. Specifically, taking $H \simeq 1/(2 t)$, as appropriate for a radiation-dominated epoch, Eq.\eqref{eq:scalarH} has a well-known solution in terms of Bessel functions, which takes the approximate  form
\be \label{eq:barthetaRD}
    \bar \theta(t) \simeq \theta_i
                    \begin{cases}   1 - \frac{(m t)^2}{5} & \quad \text{for} \qquad m / H \ll 1 \\
                                    \mathcal{Z} \frac{\sin (m t + \pi / 8)}{(m t)^{3/4}} & \quad \text{for} \qquad m / H \gg 1
                    \end{cases} \ ,
\ee
with $\mathcal{Z} \equiv \frac{2^{3/4} \Gamma\left( \frac{5}{4} \right)}{\pi^{1/2}} \simeq 0.9$. If the axion is stable on cosmological timescales, its current abundance is given by
\begin{align}
\label{eq:ALP_dark_matter_prediction}
     \Omega_{\rm ALP} \simeq 0.25 \, \theta_i^2  \sqrt{ \frac{m}{0.2 \eV}}
        \left(\frac{f}{10^{12} \GeV} \right)^2 \left( \frac{106.75}{g_*(T_{\rm osc})}\right)^{1/4} ,
\end{align}
where $T_{\rm osc}$ is the temperature at which $m \sim H$ and the axion field starts oscillating, and a standard cosmology has been assumed since that temperature.~\footnote{In \ref{sec:QCD_axion_and_EWPT} we will comment on the effect of  a super-cooled phase transition on this prediction (see Eq.\eqref{eq:ALP_DM_with_super-colling}).} Inverting Eq.\eqref{eq:ALP_dark_matter_prediction} gives the mass as a function of the decay constant (and $\theta_i$)
\begin{align}
\label{eq:massFromf}
    m \approx 0.2 \eV \, r_{\rm dm, 0}^2 \, \left( 10^{12} \GeV \over \theta_i f \right)^4 \ ,
\end{align}
where $r_{\rm dm, 0} \equiv \Omega_{\rm ALP}/\Omega_{\rm CDM}$ and $\Omega_{\rm CDM} \simeq 0.25$.
ALPs produced via misalignment can therefore account for all of the dark matter today ($r_{\rm dm, 0}=1$) or a fraction of it ($r_{\rm dm, 0}<1$). 
Values of $f$, $m$ and $\theta_i$ that would lead to overproduction ($r_{\rm dm, 0}>1$) are also possible provided the ALP decays sufficiently early. It may, for example, decay to photons via an electromagnetic coupling
\begin{align}
\label{eq:axion_photon_coupling}
    \mathcal{L} \supset {1\ \over 4} g_{a\gamma \gamma} a F \tilde{F}
    \qquad \text{where} \qquad a(x) \equiv f \theta(x)
    \qquad \text{and} \qquad g_{a\gamma \gamma} \equiv {\alpha_{\rm em}\over 2 \pi f}c_{a,\gamma} \ ,
\end{align}
which is also the most studied SM-axion interaction for experimental purposes (see e.g. \cite{AxionLimits} for a wide range of current and future constraints). Above, $\alpha_{\rm em} \approx 1/137$ and $c_{a,\gamma}$ is a UV sensitive coefficient. This gives a lifetime
\begin{align}
    \tau = { 64 \pi \over g_{a\gamma \gamma }^2 m^3  } \approx \tau_0 \left(80 \keV \over m \right)^3\left(f/ c_{a,\gamma}  \over 4 \cdot  10^{10}\GeV\right)^2 \ ,
\end{align}
where $\tau_0 \sim 3 \cdot  10^{17}s$ is the age of the universe, and we have normalised to the values of $m,f$ resulting in $r_{\rm dm,0}=1$ for $\theta_i=1$ that would be decaying at the current epoch. 

We begin in \ref{sec:gammaeq} evaluating the pressure experienced by relativistic thin walls in the presence of either a frozen or an oscillating axion background, and discuss the necessary requirements for this effect to bring the expanding walls to a regime of (nearly) constant velocity. In \ref{sec:tmunu} we discuss the effect of the axion-wall interactions on the background axion field and comment on the potential impact of this effect on the gravitational wave signature of a first order phase transition. In \ref{sec:Relativistic_axions} we explore the possibility that a small component of relativistic axions is present at the current epoch, as a result of axion-wall interactions in the early universe. Finally, in \ref{sec:QCD_axion_and_EWPT} we comment on the potential effect of ALPs on the dynamics of the electroweak phase transition.

In the remainder of this section we will use a $*$ subscript to denote quantities evaluated at the time at which relativistic bubble or domain walls interact with an axion background. Thus, $T_*$, $H_*$ and $t_*$ refer to the SM plasma temperature, Hubble rate and cosmological time parameter respectively at the relevant epoch. For a first order phase transition, $t_* \sim t_n$ (the time at which bubbles nucleate), whereas for a domain wall network $t_*$ would refer to the time at which the network collapses and the domain walls reach relativistic velocities.

\subsection{Terminal velocity}
\label{sec:gammaeq}

We can obtain the pressure experienced by relativistic thin walls during a radiation dominated epoch by substituting Eq.\eqref{eq:barthetaRD} into our expression Eq.\eqref{eq:P2}. First, during the frozen regime:
\be
    \mathcal{P} (t_*) \simeq \frac{\gamma^2}{25} \left( \frac{\Delta f^2}{f^2} \right)^2 \rho_i \left( \frac{m}{H_*} \right)^2
    \qquad \qquad \textrm{($ m/H_* \ll 1$)} \ , \label{eq:P_RD_frozen}
\ee
where $\rho_{\theta, i} \equiv \frac{1}{2} \theta_i^2 f^2 m^2$ is the initial energy density in the axion field, which stays approximately constant while $m / H_* \ll 1$, and we have neglected the first term in Eq.\eqref{eq:P2} (subdominant in the relativistic limit). On the other hand, once the axion field starts oscillating, we find
\begin{align}
    \mathcal{P}(t_*)  & \simeq \gamma^2 \left( \frac{\Delta f^2}{f^2} \right)^2 \rho_\theta(t_*) \cos^2 (m t_* + \pi / 8) \qquad \qquad \textrm{($m/H_* \gg 1$)} \label{eq:P_RD_osc} \\
                    & \simeq \frac{\gamma^2}{2} \left( \frac{\Delta f^2}{f^2} \right)^2 \rho_\theta(t_*) \ , \label{eq:P_RD_osc_avg} \\[-10pt]
                    & \, \uparrow \nonumber \\[-5pt]
                    & \textrm{\footnotesize{average}}  \nonumber
\end{align}
where in the last step we have averaged over timescales $\gg m^{-1}$. Here, $\rho_\theta(t_*)$ refers to the axion energy density in the oscillating regime, namely $\rho_\theta(t_*) \simeq \frac{\mathcal{Z}^2 \rho_{\theta, i}}{(m t_*)^{3/2}}$. If the universe was matter- rather than radiation-dominated Eq.\eqref{eq:P_RD_frozen}-\eqref{eq:P_RD_osc_avg} are only modified by overall $\mathcal{O} (1)$ numerical factors.

As discussed in \ref{sec:qm}, the $\gamma^2$-enhanced pressure on relativistic wall-like objects is only present for $\gamma$-factors in the kinematic regime
\be \label{eq:inter}
    1 \ll \gamma \ll (mL)^{-1} \ ,
\ee
with $L$ the thickness of the wall (in its own rest frame). For this kinematic regime to be accessible, one must therefore have $m L \ll 1$. For bubble walls formed during a first order phase transition at finite temperature, the natural expectation is that $L \sim T_*^{-1}$ in the absence of any significant supercooling. 
On the other hand, a process of vacuum decay that takes place in a hidden sector that is thermally decoupled from the SM could feature $L \ll T_*^{-1}$. Similarly, a network of domain walls typically collapse much later than the epoch at which they were formed, in which case the wall thickness can be orders of magnitude smaller than $T_*^{-1}$. 
Unrelatedly, dark matter ALP masses can range from $m \sim 10^{-22} \ \text{eV}$ for $f \sim 10^{17} \ \text{GeV}$ to $m \sim 10^8 \ \text{GeV}$ for $f \sim m$ (assuming $\theta_i = \mathcal{O} (1)$), and a much wider range of parameters is possible if we relax the assumption that the axion makes the dark matter. The dimensionless ratio $m L$ can thus be naturally tiny during much of the early universe in many ALP extensions of the SM.

Although we are considering small changes in the axion decay constant, $\gamma$-factors in the early universe can be large enough that the pressure exerted by a cosmological axion background could neutralize the difference in energy densities at either side of the wall. When this is the case, wall-like defects will asymptote to a trajectory of nearly constant velocity, with an equilibrium $\gamma$-factor given by
\be \label{eq:gammaeq}
    \mathcal{P} (t_*) \simeq \Delta V \qquad \Rightarrow \qquad
    \gamma_\eq \simeq \frac{\sqrt{2 \Delta V / \rho_\theta (t_*)}}{ | \Delta f^2 / f^2 | }  \ ,
\ee
where we have used the expression for $\mathcal{P}$ in Eq.\eqref{eq:P_RD_osc_avg} for illustration. It is convenient to parameterize the difference in vacuum energies in terms of the total energy density at the relevant epoch, i.e.~$\Delta V \equiv \alpha \rho_\text{total}(t_*)$. Values of $\alpha$ between $10^{-3} - 10^{-1}$ may lead to observable gravitational wave signals, depending on the characteristic frequency. Eq.\eqref{eq:gammaeq} can then be conveniently written as
\begin{align}
    \gamma_\text{eq}
        & \simeq \frac{\sqrt{2 \alpha}}{|\Delta f^2 / f^2|} \left( \frac{\rho_\text{total} (t_*)}{\rho_\theta (t_*)} \right)^{1/2} \\
        & \simeq \frac{1}{\sqrt{r_{\text{dm},*}}} \frac{\sqrt{2 \alpha}}{|\Delta f^2 / f^2|} \left( \frac{\Omega_\text{rad}}{\Omega_\text{dm}} \right)^{1/2} \left( \frac{T_*}{T_0} \right)^{1/2} \left( \frac{\gstars (T_*)}{\gstars (T_0)} \right)^{1/6} \ ,
\end{align}
where $r_{\text{dm}, *} \equiv \rho_\theta (t_*) / \rho_\text{dm} (t_*)$ is the ratio of axion to the would-be dark matter energy density in the early universe. If the axion is stable on cosmological timescales then $r_{\text{dm}, *} \leq 1$ (assuming a standard cosmological history), but one could have $r_{\text{dm}, *} \gg 1$ if it decays sufficiently early. Plugging in some numbers, for illustration,
\be \label{eq:gammaeq_ex}
    \gamma_\eq \simeq \frac{2 \cdot 10^9}{\sqrt{r_{\text{dm}, *}}}
        \left( \frac{10^{-2}}{ | \Delta f^2 / f^2 | } \right) \left( \frac{\alpha}{0.1} \right)^{1/2} \left( \frac{T_*}{10^6 \ \text{GeV}} \right)^{1/2} \left( \frac{\gstars (T_*)}{106.75} \right)^{1/6} \ .
\ee
For this choice of parameters, the radius of a vacuum bubble at the onset of this equilibrium regime, relative to the overall Hubble radius, is given by
\be
    \frac{R_\eq}{H_*^{-1}} \sim \frac{\gamma_\eq R_n}{H_*^{-1}} \sim 10^{-3} \ll 1 \ ,
\ee
where $R_n$ refers to the bubble radius at nucleation and in the last step we have set $R_n \sim T_*^{-1}$, as would be natural for a `vanilla' thermal transition. Thus, bubble walls can reach this equilibrium regime when they are still small compared to their final radius at collision.

\begin{figure}
    \centering
    \includegraphics[scale=0.5]{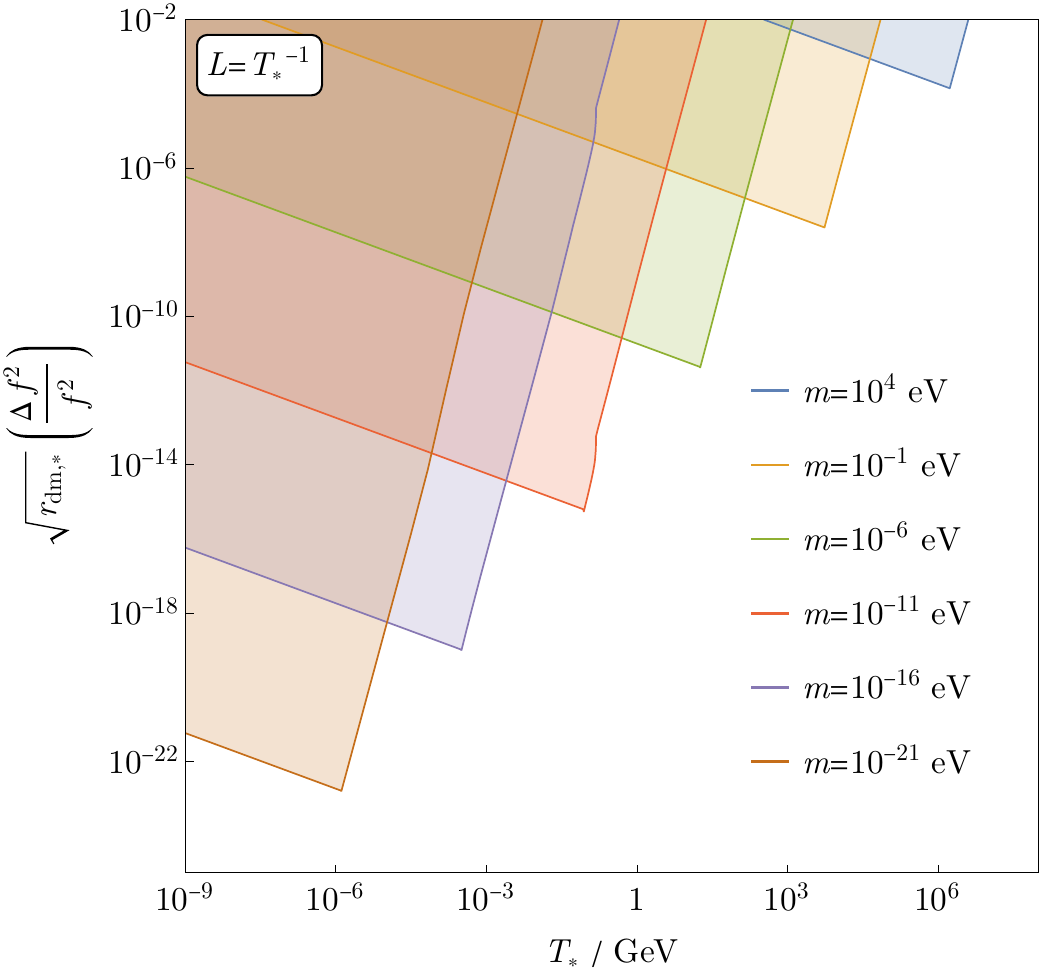}
    \caption{Colored contours correspond to values of $r_{\text{dm}, *}^{1/2} \left( \frac{\Delta f^2}{f^2} \right)$ for which the bubble walls of a cosmological phase transition reach an equilibrium velocity, as a function of the temperature of the thermal plasma at which the transition takes place. The contours are obtained by requiring that $\gamma_\eq \leq \min \left\{ (mL)^{-1}, (H_* R_n)^{-1} \right\}$. Each color corresponds to a different ALP mass, as indicated. For illustration, we take $\alpha = 0.1$, $R_n = L$, and $L=T_*^{-1}$.
    } 
    \label{fig:plots}
\end{figure}

Fig.~\ref{fig:plots} shows the region of parameter space where pressure from a cosmological axion background can halt the acceleration of the expanding bubble walls. The colored contours correspond to regions of parameter space where the walls reach a terminal velocity, for different values of the ALP mass as indicated in the figure. The various contours are obtained by requiring that $\gamma_\text{eq}$ lies below $\min \left\{ (mL)^{-1}, (H_* R_n)^{-1} \right\}$, and $\gamma_\text{eq}$ has been computed by solving $\mathcal{P} (t_*) = \Delta V$ with $\mathcal{P} (t_*)$ given by Eq.\eqref{eq:P_RD_frozen} and Eq.\eqref{eq:P_RD_osc_avg} in the frozen and oscillating regimes respectively. For illustration, we have chosen $R_n = L = T_*^{-1}$, as expected for a standard thermal transition. The requirement that $\gamma_\eq \lesssim (mL)^{-1}$ dominates for temperatures such that $H_* \lesssim m$ (oscillating regime), whereas $\gamma_\text{eq} \lesssim (H_* R_n)^{-1}$ is the dominant constraint when $H_* \gtrsim m$ (frozen regime). If the phase transition was supercooled, then $v \sim R_n^{-1} \sim L^{-1} \gg T_*$, and the colored contours of Fig.~\ref{fig:plots} are displaced down by the corresponding supercooling factor $\sim T_* / v \ll 1$, further opening the region of parameter space where axion pressure can affect the evolution of the bubble walls.

An important consideration concerns the `expected' size of the ratio $\Delta f^2 / f^2$. In the presence of interactions such as Eq.\eqref{eq:dim6}, and assuming a standard thermal phase transition, one expects $\langle \phi \rangle \sim T_*$ and therefore
\be \label{eq:df2_exp}
    \frac{\Delta f^2}{f^2} \sim \frac{\kappa T_*^2}{f^2} \ .
\ee
Within the UV-completion discussed in \ref{sec:setup}, avoiding fine-tuning in the scalar potential for the order parameter seems to require $\kappa \lesssim m_\phi^2 / m_\rho^2$, which must be $\lesssim 1$ but may be $\gg v^2 /f^2$ in the presence of a large hierarchy of quartic couplings $\lambda_\Phi \ll \lambda_\phi$, making the axion radial mode parametrically lighter than the PQ-breaking scale -- a logical possibility albeit not necessarily generic. 
For example,
\be \label{eq:df2f2_ex}
    r_{\text{dm},*}^{1/2} \left( \frac{\Delta f^2}{f^2} \right) \sim 10^{-17}  \kappa \, \theta_i \left( \frac{m}{0.1 \ \eV} \right)^{1/4} \left( \frac{T_*}{5 \cdot 10^3 \ \GeV} \right)^2 \left( \frac{10^{12} \ \GeV}{f} \right) \ ,
\ee
where we have normalized the various parameters taking into consideration that an axion with mass $m \sim 0.1 \ \eV$ and decay constant $f \sim 10^{12} \ \GeV$ would naturally account for all of the dark matter when $\theta_i \sim 1$. Eq.\eqref{eq:df2f2_ex} is to be compared with the orange contour in Fig.~\ref{fig:plots}. If $\kappa \lesssim 1$, the right-hand-side above is clearly too small for axion friction to halt the acceleration of the bubble walls. Notice that for the ALP misalignment scenario discussed here, friction becomes more important as $f$ is decreased for fixed mass, despite underproducing dark matter.
In any case, we generally require $\kappa \gg 1$, which as discussed in \ref{sec:setup} would introduce a commensurate amount of fine-tuning into the scalar potential of the order parameter.
Other UV-completions of Eq.\eqref{eq:dim6} where this fine-tuning is less severe (and therefore axion friction more naturally important for standard thermal transitions) may be possible -- an important topic that we leave for future investigation.

\begin{figure}
    \centering
    \includegraphics[scale=0.5]{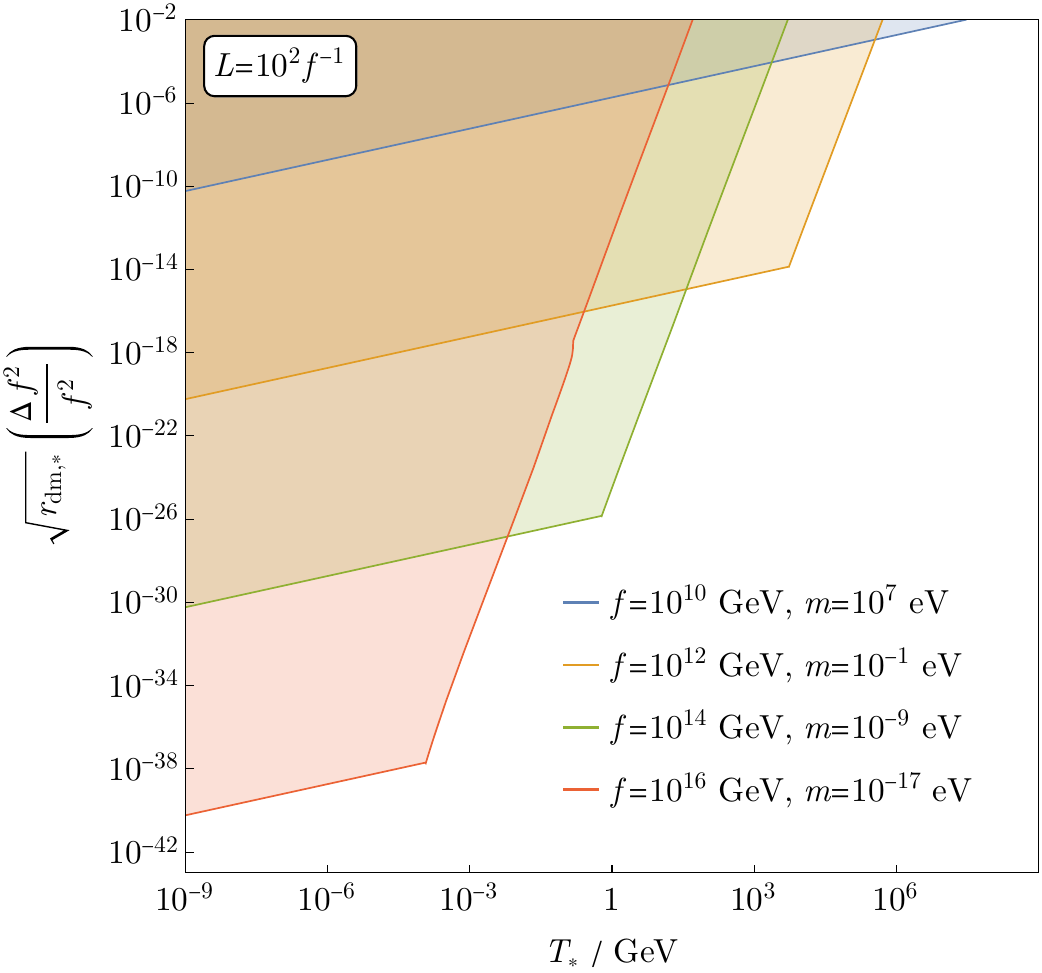}
    \caption{The various colored contours are obtained as described in the caption of Fig.~\ref{fig:plots}. Each color corresponds to a different ALP mass and decay constant, as indicated, such that the correct dark matter relic density is obtained via misalignment with $\theta_i \sim 1$. For illustration, we take $\alpha = 0.1$, $R_n = L = v^{-1}$ with $v = 10^{-2} f$.} 
    \label{fig:cold}
\end{figure}

A qualitatively different possibility concerns a phase transition -- or, rather, vacuum decay process -- that takes place within a cold hidden sector that is decoupled from the SM. 
In this case, the corresponding bubble nucleation rate per unit volume, $\Gamma_n$, would stay constant (independent of the temperature of the thermal plasma), but the relative nucleation probability $\Gamma / H^4$, would still increase as $T$ drops during the early universe. If $\Gamma / H_*^4 \sim 1$ at some temperature $T_*$ then vacuum decay could take place within this cold sector. In this case, the characteristic scale that sets the size and thickness of the bubble walls is independent of the temperature of the thermal plasma and can be $v \sim R_n^{-1} \sim L^{-1} \gg T_*$. This significantly increases the expected size of the ratio $\Delta f^2 / f^2$ compared to Eq.\eqref{eq:df2_exp}, and opens the region of parameter space where axion pressure can drive bubble wall motion to a regime of constant velocity in a way that is fully natural (in the sense discussed in \ref{sec:setup}).
Fig.~\ref{fig:cold} shows the relevant region of parameter space for $v = 10^{-2} f$, for various values of $f$ and the axion mass $m$ such that the dark matter relic density is obtained via misalignment with $\theta_i \sim 1$ (recall Eq.\eqref{eq:ALP_dark_matter_prediction}). For comparison, $\Delta f^2 / f^2 \sim \kappa \, 10^{-4}$ in this scenario, which can fall well within the colored contours of Fig.~\ref{fig:cold} even for tiny values of $\kappa$.

Overall, sufficiently large pressure due to a cosmological axion background appears more natural in the context of phase transitions that take place in a decoupled hidden sector with a characteristic energy scale that exceeds the temperature of the thermal plasma.

\subsection{Energy budget and relativistic axion shells}
\label{sec:tmunu}

If a relativistic bubble or domain wall reaches a terminal velocity as a result of axion-wall interactions, the energy released in the phase transition no longer goes into accelerating the interface, but rather it creates a disturbance in the axion field. Whether this disturbance survives until later times depends on its interactions with the surrounding medium, a discussion we defer to \ref{sec:Relativistic_axions}. Here, we first focus on describing the disturbance produced in the axion field, and understanding the properties of the corresponding energy-momentum tensor. For simplicity, we focus on the case of an oscillating axion background, and assume the wall can be treated as planar. In this case, the solution for the axion field can be written in terms of incident, reflected and transmitted components, as given in Eq.\eqref{eq:theta_sol}.

In particular, the reflected contribution to the axion field can be written as
\be \label{eq:thetar}
    \theta_r (t,z) \simeq r_\theta \bar \theta (t_*) \cos (\omega_r \Delta t - k_r z + \delta)
\ee
for $z \in [z_0(t), z_0(t) + \Delta z]$, and vanishing outside this region, with
\be \label{eq:deltaz}
    \Delta z = (1-v) \Delta t \simeq \frac{\Delta t}{2 \gamma^2} \ .
\ee
$\Delta z$ is the thickness of the reflected axion `shell', which builds up as the bubbles expand, and $\Delta t = t -t_n \geq 0$ as usual. In Eq.\eqref{eq:thetar}, $\bar \theta (t_*)$ refers to the amplitude of the axion background at the time of the phase transition, e.g.~$\bar \theta (t_*) \simeq \mathcal{Z} \theta_i  / (m t_*)^{3/4}$ for a radiation-dominated universe (remember Eq.\eqref{eq:barthetaRD}), and $\delta$ a constant that in general is non-zero. $\omega_r$ and $k_r$ are the energy and momentum of reflected axions in the axion rest frame (that is, the frame where the wall moves with Lorentz factor $\gamma$), and they are given by
\be \label{eq:wr}
    \omega_r = (1+v^2) \gamma^2 m \simeq 2 \gamma^2 m
    \quad \qquad \text{and} \qquad \quad
    k_r = 2 v \gamma^2 m \simeq 2 \gamma^2 m \ ,
\ee
in the relativistic limit. In particular, notice that the $\gamma$-factor of the reflected axion component is enhanced by an extra factor of $\gamma$, i.e.~
\be
    \gamma_\text{shell} \equiv \frac{\omega_r}{m} \simeq 2 \gamma^2 \gg \gamma \ .
\ee

The leading contribution to the energy-momentum tensor of the axion field arises from this reflected shell of axions. All diagonal components of $T_\text{shell}^{\mu \nu}$, as well as $T_\text{shell}^{03} = T_\text{shell}^{3 0}$, are non-zero in the region of thickness $\Delta z$ in front of the wall. Let us define the spatial average of each component as
\be
    \langle T_\text{shell}^{\mu \nu} \rangle \equiv \frac{\int dz \, T_\text{shell}^{\mu \nu}}{\Delta z} \ .
\ee
With this (standard) definition, we have
\be \label{eq:TR00}
    \langle T_\text{shell}^{00} \rangle \simeq  \langle T_\text{shell}^{33} \rangle
    \simeq \langle T_\text{shell}^{03} \rangle \simeq 4 \gamma^4 r_\theta^2 \rho_\theta (t_*) \ ,
\ee
and $r_\theta^2 \simeq \left( \Delta f^2 / 2 f^2\right)^2$ (recall Eq.\eqref{eq:R}). $\rho_\theta (t_*)\equiv \frac{1}{2} f^2 m^2 \bar \theta (t_*)^2$ is the energy density in the background axion field at $t \simeq t_*$, and $\langle T_\text{shell}^{11} \rangle = \langle T_\text{shell}^{22} \rangle = \mathcal{O} (\gamma^0)$. The energy density in this reflected axion shell is highly localized over a small distance $\Delta z$ in front of the expanding wall. Per unit area of the wall, the corresponding energy density is given by
\begin{align} \label{eq:T00dz}
    \langle T_\text{shell}^{00} \rangle \Delta z
    & \simeq 2 \gamma^2 r_\theta^2 \rho_\theta (t_*) \Delta t \\
    & \simeq \Delta V \Delta t \label{eq:dVdt}
    \ ,
\end{align}
and to arrive at the last step we have substituted $\gamma = \gamma_\text{eq}$, as given in Eq.\eqref{eq:gammaeq}. Eq.\eqref{eq:dVdt} is precisely the energy per unit area that is released as the wall expands over a timescale $\Delta t$. Thus, when the pressure created by a cosmological axion background stops the acceleration of the bubble walls, all of the energy released as the bubbles grow goes into forming this shell of relativistic axions.

For comparison, the (similarly) averaged energy-momentum tensor of a relativistic planar domain wall moving along the $z$-direction has non-zero components
\be
    \langle T^{00}_\text{dw} \rangle \simeq \langle T^{33}_\text{dw} \rangle \simeq \langle T^{03}_\text{dw} \rangle \simeq \frac{\gamma^2_\text{dw} \sigma_\text{dw}}{L_\text{dw}} \ ,
\ee
whereas $T^{11}_\text{dw} = T^{22}_\text{dw} = \mathcal{O} \left( \sigma_\text{dw} / L_\text{dw} \right)$. Here, $\sigma_\text{dw}$, $L_\text{dw}$ and $\gamma_\text{dw}$ refer to the tension, thickness, and $\gamma$-factor of the domain wall respectively. Thus, to leading order in $\gamma \gg 1$, the energy-momentum tensor of the reflected axion shell has the same form as that of a relativistic, planar domain wall with $\gamma_\text{dw} \simeq \gamma$, thickness $L_\text{dw} \simeq \Delta z$ (as given in Eq.\eqref{eq:deltaz}) and effective tension $\sigma_\text{dw} \simeq 2 r_\theta^2 \rho_\theta (t_*) \Delta t $. To first approximation, it therefore seems plausible that the gravitational wave radiation that is emitted when these shells collide will be well-described within the so-called `envelope approximation'~\cite{Kosowsky:1991ua,Kosowsky:1992rz,Kosowsky:1992vn,Kamionkowski:1993fg}, which assumes that the energy released in the phase transition accumulates in a thin shell surrounding the expanding bubbles. Nonetheless, there are two important differences that may lead to the corresponding stochastic gravitational wave background to differ from that of standard bubble wall collisions. First, unlike in traditional scenarios, the energy in these axion shells may not dissipate quickly after the shells meet. Indeed, comparing the mean-free-path of the highly relativistic reflected axions to the shell thickness, we find
\be \label{eq:ratio}
    \frac{\lambda_\text{mfp}}{\Delta z} \sim \frac{10^{11}}{r_\text{dm}^3 \eta^2} \left( \frac{m}{1 \ \text{eV}} \right)^3 \left( \frac{\alpha}{0.1} \right)^2 \left( \frac{10^4 \ \text{GeV}}{T_*} \right) \left( \frac{106.75}{g_* (T_*)} \right) \left( \frac{10^{-4}}{\Delta f^2 / f^2} \right)^6 \ .
\ee
This ratio can clearly be $\ggg 1$ in much of the relevant parameter space, but obviously there is a strong dependence on the various underlying parameters. We have estimated the mean-free-path as $\lambda_\text{mfp} \sim (n_\text{dr} \sigma)^{-1}$, with $\sigma$ the total axion annihilation cross section at high energies $\sqrt{s} \simeq 2 \omega_r$. Since $\sqrt{s}$ is well above the cutoff of the EFT, one must use the full UV-completion described in \ref{sec:setup} to compute $\sigma$. For the parameters highlighted in Eq.\eqref{eq:ratio} the inclusive cross section is dominated by the $\theta \theta \rightarrow \phi \phi$ channel, and thus we take $\sigma \sim \eta^2/(8 \pi s)$ in this example. This lack of immediate dissipation could lead to this source of gravitational waves lasting during timescales $\gg H_*^{-1}$, in turn enhancing the corresponding signal (in terms of the usual parametrization, see e.g.~\cite{Caprini:2015zlo,Caprini:2019egz}, this would correspond to a parameter $\beta \gg 1$.) Second, the presence of a component of free-streaming particles at the time of gravitational wave production could further affect the shape of the corresponding signal at low frequencies. This has been discussed in the context of a homogeneous free-streaming component of the total energy density at the time of gravitational wave emission in~\cite{Hook:2020phx,Loverde:2022wih}. 
In the scenario described here, the free-streaming component in instead highly localized, and could lead to additional features on the scale of the shell thickness. Overall, the gravitational wave signal stemming from collisions of relativistic particle shells is poised to feature significant differences from those of commonly discussed sources and it is an interesting direction for future investigation. 

We have focused on the properties of the axion energy-momentum tensor during the oscillating regime, but an analogous effect takes place if the axion is frozen. In that case, a numerical study of Eq.\eqref{eq:theta1_wall} or \eqref{eq:theta1_bubble} shows that
a sharply sloped disturbance localized near the expanding wall rises out of the near constant background axion field  -- an `axion cliff'. In the following section we will simply interpret these as particles produced with energy $\omega_r$, and leave a more detailed analysis of their $T_{\mu\nu}$ and evolution to future work.

In this subsection, we have neglected the effect of interactions between the axions and a potential thermal population of $\phi$ particles. 
This is certainly appropriate when the phase transition in question takes place in a cold hidden sector decoupled from the thermal plasma (see Fig.~\ref{fig:cold} and surrounding discussion), but in a standard thermal transition, interactions of the form $\phi \, \theta  \rightarrow \phi \, \theta$ might dissipate energy and preclude the shells/cliffs from forming, as discussed more quantitatively below.

\subsection{Relativistic axion background today}
\label{sec:Relativistic_axions}

We now discuss the interesting possibility that the fraction of dark matter axions boosted by the wall in the early universe survives unperturbed until the present day, forming a relic relativistic background of a qualitatively new type. 
For this, they must scatter inefficiently with each other (discussed in the previous section around Eq.(\ref{eq:ratio})) as well as the surrounding medium, throughout the intervening cosmic history. Otherwise, they might be absorbed by the medium, or lose most their kinetic energy, ultimately forming a tiny fraction $\mathcal{R}$ of the DM today. 
In the case of a thermal PT, scattering with the thermal population of excitations of the order parameter $\phi$, through the same coupling Eq.\eqref{eq:dim6} responsible for the boosted axions to start with, can in general be quite efficient. 
To see this in more depth, and quantitatively follow the fate of a reflected axion,
we will now proceed by highlighting an example `thermal' transition. At the end, we will further comment on `cold' transitions, which are less constraining.

Similar to \ref{sec:gammaeq}, we characterize `thermal' by setting all scales relevant for the PT to the temperature $v,L^{-1},R_n^{-1}\sim T_*$. 
We will consider $\alpha = 0.1$, meaning the latent heat released $\Delta V$ is $10\%$ of the SM bath energy at the time.
For the background ALP field,
for concreteness  we will take it to give the observed dark matter ($r_{\rm dm, 0}=1$) by the misalignment mechanism \eqref{eq:ALP_dark_matter_prediction}, fixing the relation between mass and decay constant \eqref{eq:massFromf}
by taking $\theta_i = 1$ for simplicity (and no good physical reason). 
Normalizing to a particular benchmark point that falls into the oscillating regime,
after reflection axions have an energy given by Eq.\eqref{eq:wr}, with $\gamma = \gamma_\text{eq}$:
 \begin{align}
 \label{eq:omega_r}
     \omega_r \approx 4 \cdot  10^{14}\GeV  \left(\alpha \over 0.1\right)\left(  \frac{10^{12} \GeV}{f}  \right)^4 \left( \frac{3\cdot 10^{3} \GeV}{T_*} \right)^3 \left( m_\rho \over 10^6 \GeV  \right)^4 \left(0.1 \over \eta \right)^2\ , 
 \end{align}
 which leads to a centre-of-momentum energy of $\sqrt{s}/2 \approx \sqrt{2 \omega_r T_*}/2 \sim 8 \cdot  10^8 \GeV \gg m_\rho$. Thus $a\phi \rightarrow a\phi$ scattering must be computed in the UV theory, which we take to be Eq.\eqref{eq:dim4} for simplicity. The interaction rate $\Gamma_{\phi a}$ is given by the cross-section and number density of $\phi$, which we take to be
 \begin{align}
 \label{eq:sigma_n_phi}
     \sigma_{a\phi } \approx {\eta^2 \over 64 \pi^2 s} \ , \quad n_\phi \approx {\zeta(3) \over \pi^2} T_*^3 \ ,
 \end{align}
 ignoring any potential over-densities in $n_\phi$ that may form  during bubble expansion.
 Comparing to the expansion rate at the time of the phase transition,
 \begin{align}
     \frac{\Gamma_{a\phi}}{H} \simeq \frac{\sigma_{a\phi } \, n_\phi}{H} \approx 0.01 \left(0.1 \over \alpha \right) \left( \frac{f}{10^{12} \GeV} \right)^4 \left( \frac{T_*}{3 \cdot 10^3 \GeV} \right)^3 \left( 10^6 \GeV \over m_\rho \right)^4 \left(\eta \over 0.1 \right)^4 \ .
 \end{align}
While our particular benchmark point is safe, we see that in other cases scattering can easily be efficient. 
The reader might protest that, while \eqref{eq:sigma_n_phi} are true, $\Gamma_{\rm a \phi}/H$ will grow with time and might become efficient at some $T<T_*$. However, the mass of the order parameter $\phi$ will naturally  be close to the scale of the phase transition, such that $n_\phi$ and therefore $\Gamma_{a\phi}$ would start to be Boltzmann suppressed thereafter. 
For non-thermal vacuum transitions this population of $\phi $ particles will not be present.

\begin{figure}
    \centering
    \includegraphics[width=\textwidth]{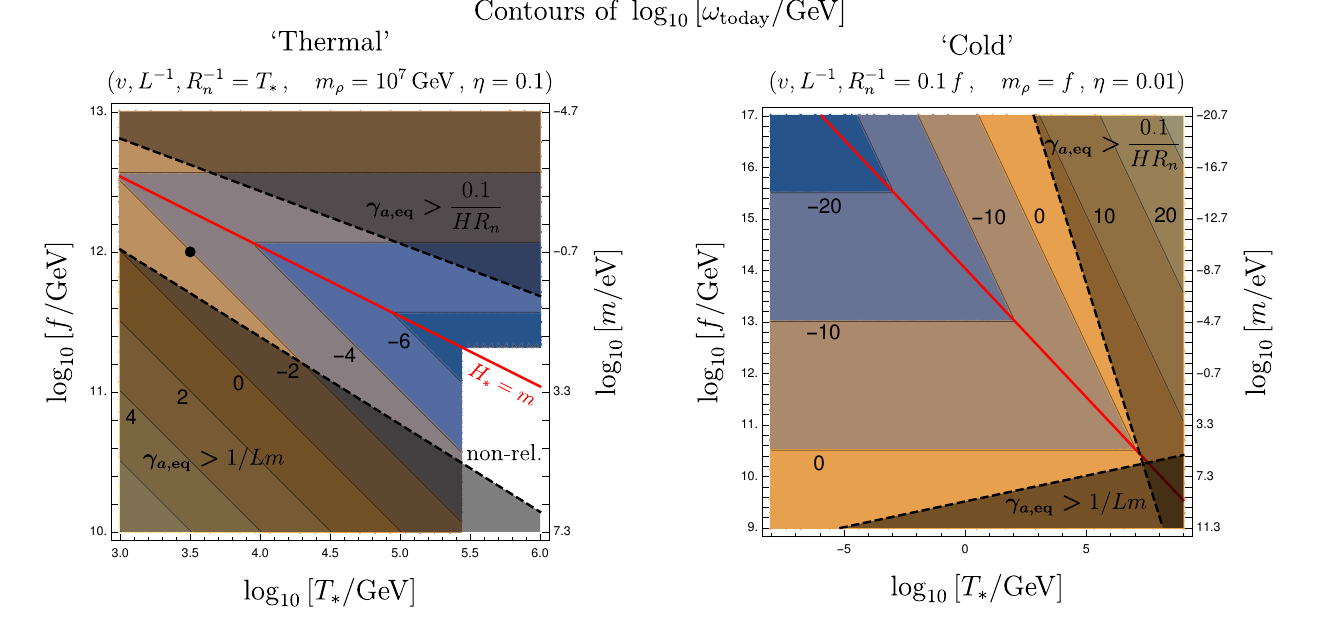}
    \caption{Dark matter axions kicked by a phase transition in the early universe can potentially survive till today as relativistic relics, though this is highly sensitive to many  parameters. We show here contours of the present day energy of the relic axions for two example scenarios with $\alpha = 0.1$ (which saturates the limit $\Delta N_{\rm eff}\lesssim 0.3$), and the ALP DM field is fixed by the misalignment mechanism with $\theta_i=1$ and $\Omega_{\rm ALP} \approx 0.25$. The red line ($H=m$) separates regimes when the axion is frozen (top) and oscillating (bottom) during the PT. The two shaded regions denote points where equilibrium due to axion friction is inconsistent (bottom), or too late (top). \textbf{Left:} `Thermal transition' -- all scales $v,L,R_n$ are set by $T_*$. The $PQ$ radial mode is taken relatively light, $m_\rho = 10^7\GeV$, and $\eta = 0.1$. The black dot corresponds to the benchmark point explored in text. In the white region reflected axions lose their kinetic energy just by redshift, and are a tiny fraction of the dark matter today. \textbf{Right:} `Cold transition', all scales are set by the UV $f$, though the transition occurs much later by tunnelling. $\eta = 0.01$.}
    \label{fig:BenchALPth}
\end{figure}

If the axions survive, they will have an energy today which is simply Eq.(\ref{eq:omega_r}) redshifted appropriately
\begin{align}
\label{eq:omega_today}
        \omega_{r,0} \approx & \; \omega_r \left( T_{\gamma,0} \over T_*\right) \left( 3.94 \over g_{*,s}(T_*) \right)^{1/3}\\
    \approx & \; 0.01 \GeV \left(\alpha \over 0.1\right) \left( 10^{12} \GeV   \over f  \right)^4 \left(  3 \cdot 10^{3} \GeV \over  T_*\right)^4 \left( 10^6 \GeV \over m_\rho \right)^4 \left( 106.75 \over g_{*,s}(T_*) \right)^{1/3}  \left(0.1 \over \eta \right)^2\ , \nonumber
\end{align}
where $ T_{\gamma,0}\approx 2.3\cdot  10^{-4}\eV$ is the CMB temperature. Again we see the strong dependence on the underlying parameters. The reader should be cautioned against trusting \eqref{eq:omega_today} for very different parameters. For values crossing into the frozen regime ($m<H_*$) the parametric dependence of equilibrium $\gamma_{\rm eq}$ (and $\omega_{r,0}$) is changed by the extra factor of $m/H_*$ (and $m^2/H^2_*$) as per \eqref{eq:P_RD_frozen}. Moreover, the consistency constraints $\gamma_{\rm eq}<1/mL$ and $\gamma_{\rm eq}\ll 1/Hr_n$ are not satisfied for every point, but  are often non-trivial in `thermal' examples where $L,R_n \sim T_*^{-1}$. 
In the left panel of Fig.~\ref{fig:BenchALPth} we show the neighbourhood of the benchmark point allowing $f$ and $T_*$ to vary. The curtain-like constraints draping the $H=m$ line open (close) as $L,R_n$ are made smaller (larger). 

In the right panel of \cref{fig:BenchALPth} we instead study examples of 'cold' vacuum transition, where all scales related to the transition are set by $f$. As mentioned in \ref{sec:gammaeq}, within the simple UV picture of \eqref{eq:axion_photon_coupling}, this is the most natural expectation, barring the usual tuning necessary for any cold transition to complete before the onset of eternal inflation. As reflected in the two panels, this case is generally far less constrained. 
In both cases, one might also worry about axions scattering off the SM plasma, for example through \eqref{eq:axion_photon_coupling}. We find this highly subdominant in all regions of Fig.~\ref{fig:BenchALPth}.

\paragraph{Dark radiation.} Assuming the axions remain relativistic long enough, they will contributed to dark radiation at CMB (or BBN) scales, conventionally measured in terms of an effective number of extra neutrino species $\Delta N_{\rm eff}$. Recall $\alpha$ was the fraction of SM  energy density released in the form of latent heat $\Delta V$ at the PT scale $T_*$, and this was efficiently transferred to the reflected axions. It is then straightforward \cite{GarciaGarcia:2022yqb} to translate $\alpha$ to  
\begin{align}
\label{eq:DeltaNeff_from_alpha}
    \Delta N_{\rm eff}
    \simeq 0.3  \left( \frac{\alpha}{0.1} \right) \left( \frac{106.75}{g_{*}(T_*)} \right)^{1/3} \ ,
\end{align}
and compare to the bound  $\Delta N_{\rm eff} < 0.28$ at $95 \%$ C.L. from (Planck 18 + BAO) \cite{Aghanim:2018eyx}. 

\paragraph{Flux today.}
We can ask what is the present day flux of these axions on Earth, 
\begin{align}
\label{eq:Axion_flux}
    \mathcal{F}_{a}
    & \simeq 65 \left(\Delta N_{\rm eff} \over 0.3 \right)\left(0.05 \GeV  \over  \omega_{r,0} \right) \cm^{-2}s^{-1} \ ,
\end{align}
which applies more broadly to any present relativistic component of monochromatic energy $\omega_{r,0}$ contributing a $\Delta N_{\rm eff}$ amount to dark radiation. For our benchmark energy, this is an underwhelming number. However, Fig.~\ref{fig:BenchALPth} shows that other points can easily lead to lower energies $\omega_{r,0}$, and therefore larger fluxes, by many orders of magnitude. Moreover, Fig.~\ref{fig:BenchALPth} and the general set-up discussed here is by no means an exhaustive -- nor necessarily the most natural -- parameter space to consider. Notice that, having fixed the total energy density in the relativistic axions, there is always a tradeoff between flux and axion energy. Experimentally, larger fluxes entail more events in a detector, but larger energy  is often easier to detect. Thus, what is observable is a function of the details of a particular experiment. Our goal at present is simply to highlight this possibility, and leave a more systematic study of the expected range of relic energies and their detection prospects to future work.

\paragraph{A comparison with other relativistic relics:}
Any new source of present day relativistic axions should be compared against other, more conservative, sources. Assuming a coupling to the SM such as \cref{eq:axion_photon_coupling}, axions are unavoidably produced inside the Sun for example, by Primakoff scattering and other processes, giving an approximate flux on Earth of \cite{Raffelt:1996wa}
\begin{align}
    \mathcal{F}_{\rm sun} \approx 200 \left( 10^{12} \GeV \over f \right)^2 \cm^{-2}s^{-1} \ ,
    \quad  \text{with} \quad \omega_{\rm sun} \sim \keV \ , 
\end{align}
where we have normalised to compare with Eq.\eqref{eq:Axion_flux}, and the average energy of the particles produced $\omega_{\rm sun}$ is set by the temperature of the sun. Solar axions are searched for in helioscopes such as CAST \cite{CAST:2015qbl} and its future upgrade IAXO \cite{IAXO:2019mpb,IAXO:2020wwp}, where they are converted into x-ray photons in the background of a strong magnetic field via the photon-axion coupling \eqref{eq:axion_photon_coupling}. The conversion probability, valid for relativistic energies $\omega \gg m$, in a detector of length $L$, is \cite{CAST:2015qbl}
\begin{align}
    P_{a\rightarrow \gamma} =  \left(g_{a\gamma \gamma}BL \over 2\right)^2 \left({ \sin\left(m^2 L / 4 \omega \right) \over m^2 L / 4 \omega }\right)^2 \ ,
\end{align}
where $L=9.26 \meter$ and $B\approx 8 \Tesla$ for the case of CAST. To have an unsuppressed conversion probability we need 
\begin{align}
    m \ll \sqrt{4 \omega L^{-1}} \approx 10^{-2} \eV \sqrt{\omega \over 1\keV} \sqrt{9.26 \meter \over L} \ .
\end{align}
which is satisfied by the example point discussed above. We dip our toes here in aspects of detection only for the fun of illustration -- by no means is this flux detectable at any foreseeable helioscope known to us. A more concrete study of relativistic reflected axions, in particular the QCD axion, for low (detectable) $f\sim 10^8 \GeV$ will be the subject of future work. 

For completeness, we should also recall that axions can be thermally produced by the SM bath in the early universe, although this is a function of reheat temperature. If this is high enough, they will thermalize for arbitrary $f$. Eventually this thermal population freezes out and their energy today unsurprisingly is close to that of CMB photons, the only difference being the number of entropy dumps in the visible sector as the temperature goes below mass thresholds. As long as the mass is light enough, the flux and energy of this relativistic relic is
\begin{align}
    \mathcal{F}_{\rm th.} \sim  0.23-1.4 \cdot  10^{12} \cm^{-2}s^{-1} \ ,
    \quad  \text{with}  \quad \omega_{\rm th.} \approx 0.78 - 1.4 \, \cdot  10^{-4}\eV \ , 
\end{align}
where the upper value corresponds to the largest contribution allowed by the dark radiation bound \cite{Aghanim:2018eyx} and the lower one is the lowest contribution for a single extra dof which was at some point in thermal equilibrium with the visible sector.

\subsection{The electroweak phase transition}
\label{sec:QCD_axion_and_EWPT}

We will now schematically examine the physics described in the previous sections in arguably one of the most concrete and constrained scenarios of interest.
As mentioned already, there is no first order phase transition in the SM. However, we will consider the frequently studied scenario where the electroweak phase transition (EWPT) is made first order with the addition of ingredients beyond the SM (see for example section 6 of \cite{Caprini:2019egz} for an overview). This has received renewed interest in recent times, as the ensuing stochastic gravitational wave background produced would be a target within the frequency range of LISA \cite{Caprini:2015zlo,2017arXiv170200786A,Caprini:2019egz,LISACosmologyWorkingGroup:2022jok}, but has a rather long history as a theory of baryogenesis \cite{Kuzmin:1985mm,Shaposhnikov:1986jp,Shaposhnikov:1987tw,Morrissey:2012db} and in the context of natural theories of the weak scale \cite{Creminelli:2001th,Craig:2014lda}. For gravitational wave detection, strong ($\alpha \gg 1$), super-cooled phase transitions are particularly of interest. Depending on the specific model, the EWPT can undergo much super-cooling, $T_* \ll v_{\rm ew} \simeq 247 \GeV$, but will certainly occur by the (symmetric phase) strong scale $\tilde{T}_c \lesssim T_c\sim 150 \MeV$ \footnote{The symmetric phase $\tilde{T}_c$ is smaller compared to the familiar broken phase $T_c$ on account of all six quarks being massless and contributing to the QCD beta function.},  which is a lower bound due to QCD catalysis \cite{Witten:1984rs}. During this period of cooling the energy density of the universe becomes dominated by vacuum energy $\Delta V$, Hubble is roughly constant at $H_{\rm ew} = \sqrt{8\pi G \Delta V/3}$, and a brief period of inflation ensues, lasting at most $N_{\rm ew}\approx \ln \left(v_{\rm ew}/\tilde{T}_c\right)\approx 7$ e-folds. Assuming instantaneous thermalization after the phase transition completes, the universe returns to be radiation dominated, with a `re-reheat' temperature of $T_{\rm rrh} \sim 0.24 \sqrt{H_{\rm ew}M_{\rm pl}}$, not to be conflated with the reheat temperature after inflation.

Focusing on the EWPT makes things more concrete, but also comes with a particularly strong source of friction already present in the SM, that is the soft emission of mass-changing gauge bosons (namely $W$s and $Z$s) from charged particles as they cross the wall. This `transition radiation' is the only process to date that has been shown to generically induce a growing pressure in the asymptotically large $\gamma$ limit: more precisely, $\mathcal{P}_{\rm tr}\sim g^3 \gamma v_{\rm ew} T_*^3/\pi^4$ per dof \cite{Bodeker:2017cim,Vanvlasselaer:2020niz,Gouttenoire:2021kjv,Azatov:2023xem}, with $g$ an electroweak gauge coupling. Summing over the SM spectrum \cite{Gouttenoire:2021kjv,Azatov:2023xem} and setting this total pressure to equal $\Delta V$ gives an equilibrium Lorentz factor
\begin{align}
\label{eq:SM_friction}
    \gamma_{\rm SM, eq} \sim 10^7 \left( \Delta V\over (100 \GeV )^4 \right)\left( 1 \GeV \over T_* \right)^3  \qquad (\text{from SM friction}) \ ,
\end{align}
which we will compare with the novel effect from axion dark matter. Despite the latter making up a tiny fraction of the energy density at the time, during the super-cooling phase this fraction grows with either one or four powers of scale factor in the oscillating and frozen regimes respectively. This, along with the faster growth $\sim \gamma^2$, suggests the effect may not always be entirely negligible.

We couple the ALP to the Higgs doublet $\mathcal H$ via the same dimension-six operator used in previous sections. The relevant part of the effective theory is thus explicitly
\begin{align}
\label{eq:Lag_EFT}
\mathcal{L}_{\rm EFT} &\supset \frac{1}{2}\left( f^2 + 2\kappa|\mathcal H|^2 \right)\left(\partial_\mu \theta \right)^2 - V(\theta) - V_{h,S}\left(|\mathcal H|,S\right) \ ,
\end{align}
where $V = \frac{1}{2} m^2 f^2 \theta^2 \, + \dots$ is the axion potential, and $V_{h,S}$ is (schematically) the potential of the Higgs doublet along with some additional fields $S$ required to make the transition first order (a single extra scalar singlet in minimal models, e.g.~\cite{Profumo:2007wc,Espinosa:2011ax}). We will not commit to a specific model of EWPT in this work. We can continue to consider Eq.\eqref{eq:dim4} as the appropriate UV completion of Eq.(\ref{eq:Lag_EFT}), with $\phi \rightarrow\mathcal H$, and thus identify the cut-off scale with the mass $m_\rho$ of the radial mode of $\Phi = (f+\rho)e^{i\theta}/\sqrt{2}$, bounded as in Eq.\eqref{eq:cutoff}. 

Before moving on to friction, one should ask what are the constraints on the dimension-six Higgs-axion coupling in Eq.\eqref{eq:Lag_EFT}, regardless of any phase transition. This was recently studied in \cite{Bauer:2022rwf}, where the strongest experimental bound comes from invisible Higgs decays and it implies
\begin{align}
\label{eq:Higgs_decay_bound}
    \frac{\kappa}{f^2} \simeq \frac{\eta}{m_\rho^2} \lesssim \frac{1}{\left(10^3 \GeV \right)^2} \ , \qquad \text{(LHC)}
\end{align} 
where the first relation is specific to the UV completion \cref{eq:dim4}.
In \ref{sec:Axion_Higgs_Portal_constraints} we complement this study by roughly checking that cosmological or astrophysical probes are not more stringent.

\begin{figure}
    \centering
    \includegraphics[width=.8\textwidth]{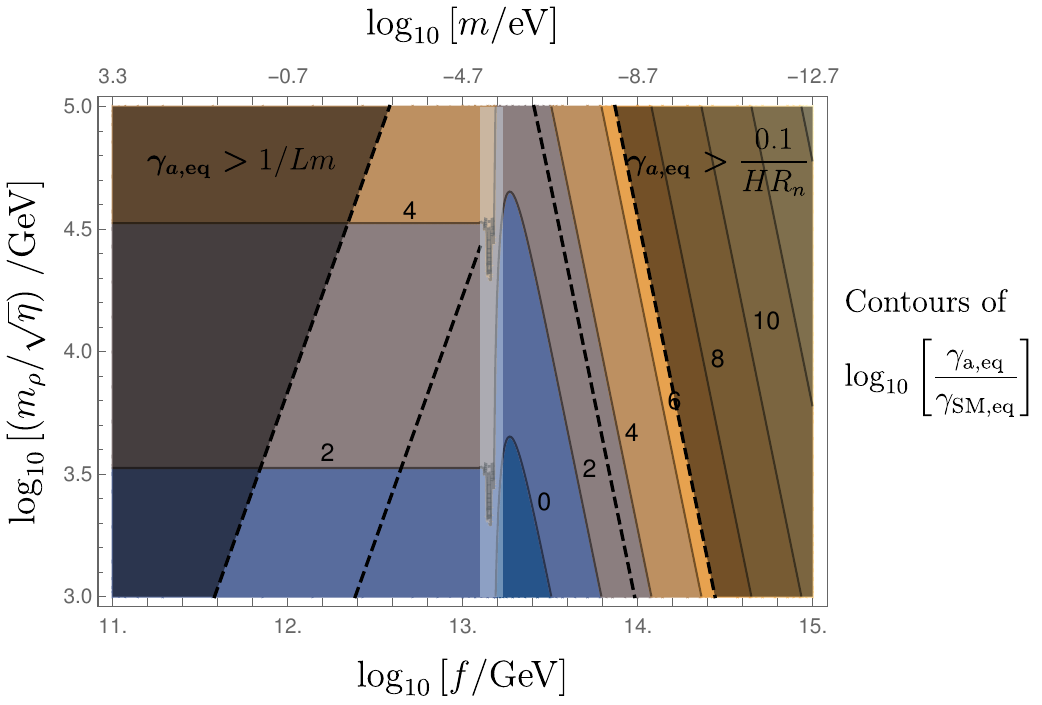}
    \caption{We study the relative importance of axion dark matter friction to that of SM transition radiation, in theories where the electroweak PT is first order and super-cooled, here as a function of axion decay constant $f$ and the effective scale $m_\rho/\sqrt{\eta}$ suppressing the axion-Higgs portal coupling. For this example, we take a vacuum energy difference $\Delta V = (100 \GeV)^4$ and a transition temperature $T_*=150 \MeV$ close to the QCD catalysis scale. The axion mass is chosen by Eq.(\ref{eq:massFromf}) with $\theta_i =1$, which gives the right present day DM in the `frozen' regime ($m \ll H_{\rm ew}$ during the PT) to the right, and a much smaller abundance in the `oscillating' regime ($m \gg H_{\rm ew}$) to the left, as explained in the text. We highlight the boundary between the two regimes around $f\sim 2\cdot  10^{13}\GeV$, where the exact solution starts to quickly oscillate, and we appropriately replace it by an average there onwards.
    $\gamma_{\rm a, eq}$ and  $\gamma_{\rm SM, eq}$ are equilibrium Lorentz factors of the bubble expansion, if the axion and SM processes were the only source of friction pressure respectively.
    For this example, the axion field dominantes only in the small dark blue region of negative contours above $f\gtrsim 10^{13}\GeV$.
    The ratio $\gamma_{\rm a, eq}/\gamma_{\rm SM, eq}$ is directly linked to the energy budget at the end of the PT according to Eq.\eqref{eq:EnergyBudget}. 
    Taking $L,R_n = v_{\rm ew}^{-1}$, the two shaded regions rule out points where equilibrium is inconsistent (left), or is too late (right). For $L, R_n = T_{*}^{-1}$, these constraints move to the inner dashed lines. 
    }
    \label{fig:EWPTandAxion}
\end{figure}

During the transition, the axion decay constant is shifted across the wall by an amount 
\begin{align}
    \Delta f^2 \simeq  \kappa v_{\rm ew}^2 \simeq \eta f^2 v_{\rm ew}^2/m_\rho^2 \ .
\end{align}
Deriving the frictional pressure from axion dark matter proceeds as before, only now in the axion EOM Eq.\eqref{eq:scalarH} Hubble remains approximately constant $H \simeq H_{\rm ew}$ coming up to the time of the phase transition. This only changes the overall $\mathcal{O}(1)$ factor in the approximate pressure formulae Eq.\eqref{eq:P_RD_frozen} and \eqref{eq:P_RD_osc}. Then, setting this equal to  $\Delta V$ gives the equilibrium $\gamma_{\rm a, eq}$ , labelled so as to distinguish it from Eq.\eqref{eq:SM_friction}. In actuality, the equilibrium expansion speed of bubbles will be the lowest of the two
\begin{align}
    \gamma_{\rm eq} = \min \left\{ \gamma_{\rm a, eq}, \gamma_{\rm SM, eq}  \right\} \ . 
\end{align}
As in \ref{sec:Relativistic_axions}, we concentrate on the scenario where the axion makes the entirety of the dark matter. 
Again, taking $\theta_i \sim 1$ for simplicity, this would determine the axion mass in terms of its decay constant $f$. 
However, the usual prediction for the abundance  Eq.\eqref{eq:ALP_dark_matter_prediction} is now distorted to account for the super-cooling period of inflation. Roughly, this amounts to an extra dilution factor 
\begin{align}
\label{eq:ALP_DM_with_super-colling}
  \begin{split}
        \Omega_{\rm ALP} \rightarrow  \Omega_{\rm ALP}' \approx \Omega_{\rm ALP}\left( T_* / T_{\rm rrh} \right)^3 \ ,  
  \end{split}
\end{align}
for those cases when the axion has already began oscillating before the start of super-cooling, i.e.~$m\gtrsim H_{\rm ew} \approx 2.4\mu\eV \sqrt{ \Delta V/ (100 \GeV )^4 }$. Notice that this is a strong effect; achieving the correct DM abundance now requires a mass $\sim\left( T_{\rm rrh} / T_* \right)^6$ larger! For significant super-cooling, we find that this increase typically comes in tension with the consistency requirement Eq.\eqref{eq:inter}, where for the super-cooled EWPT we expect $v_{\rm ew}^{-1}\gtrsim L \gtrsim T_*^{-1}$. On the other hand, if $m\ll H_{\rm ew}$ the field continues to be frozen $\bar \theta(t) \simeq \theta_i$ during this brief period of inflation, and the abundance prediction of Eq.\eqref{eq:ALP_dark_matter_prediction} remains unchanged. 

For the set of assumptions made here, we find that it is rather difficult, though not impossible, for axion dark matter to beat Eq.\eqref{eq:SM_friction} and be the dominant source of friction, as exemplified  in Fig.~\ref{fig:EWPTandAxion}, where for concreteness we took the $m(f,\theta_i=1)$ dependence of Eq.\eqref{eq:massFromf} despite the underproduction of dark matter for $f \lesssim 10^{13}\GeV$. This is reflected in the sharp increase in $\gamma_{\rm a, eq}$ below that scale. Increasing the axion mass leads to an increase in dark matter abundance like $m^{1/2}$ and therefore $\gamma_{\rm a, eq}$ decreases like $m^{-1/4}$, but at the same time the maximum $\gamma$ consistent with Eq.\eqref{eq:inter} decreases like $m^{-1}$ and the dark exclusion region to the left of Fig.~\ref{fig:EWPTandAxion} would engulf the entire region $f \lesssim 10^{13}\GeV$ before $\Omega_a = \Omega_{\rm CDM}$ is achieved. A far more significant enhancement to the axion dark matter friction would instead follow by going beyond the harmonic approximation for the axion potential discussed just after Eq.\eqref{eq:scalarH} and delay the onset of oscillations of the axion field.

In the event that the axion absorbs all or a non-insignificant fraction of the EWPT's latent heat $\Delta V$, what happens next? Quantitatively, the fraction converted to kinetic energy in the axion field is approximately
\begin{align}
\label{eq:EnergyBudget}
    {\rho_{a, \text{post-PT}} \over \Delta V} \approx \left({1 + \gamma^2_{\rm a, eq}/\gamma^2_{\rm SM , eq}} \right)^{-1} \ .
\end{align}
In the frozen case, which appears more promising at least for the EWPT, this energy 
is converted into a sharply sloped disturbance in the field, mentioned towards the end of \ref{sec:tmunu}. The fate of these `axion cliffs' depends on how efficiently they scatter with the surrounding SM plasma and the details of cliff-cliff collisions. 
Note that, unlike in \ref{sec:Relativistic_axions}, where we focused on $\alpha \ll 1$, for a super-cooled transition $\alpha \gg 1$. Therefore, 
when $\gamma_{\rm a, eq}/\gamma_{\rm SM , eq} \lesssim  1$, the cliffs actually dominate the energy density of the universe (and its expansion). In order to have a viable cosmology, certainly their energy must be efficiently transferred back to the SM sector, reheated to temperatures at least above BBN. Instead, if $\gamma_{\rm a, eq}/\gamma_{\rm SM , eq} \gtrsim 1$ only a fraction of energy goes into axion cliffs. Depending on how long they survive and if/when they turn non-relativistic, they may then have a variety of gravitational consequences (such as dark radiation, structure formation) and perhaps form a  peculiar background even today.

Finally, the reader might be wondering why we haven't focused on the QCD axion in this section, which would have made things even more concrete, e.g. fixing the relation between mass and decay constant. The answer is that it's mass in the Higgs phase is unkown. The familiar expression $m_a \approx 5.7 \cdot 10^{-6} \eV \left(10^{12} \GeV / f_a\right)$ \cite{GrillidiCortona:2015jxo} is a function of quark masses, which are zero in the unbroken phase. It still gains a mass from instanton effects, but this will be far far smaller and likely inconsequential for friction.

\section{Discussion and outlook} \label{sec:conclusions}

We have studied the interactions between two-dimensional wall-like defects and a cosmological axion background. Our main result is Eq.\eqref{eq:P_intro}, that a change in axion decay constant across the interface generates a frictional force on the moving wall that grows $\propto \gamma^2$ in the kinematic regime $1 \ll \gamma \ll (mL)^{-1}$, where $L$ is the wall thickness and $m$ the axion mass.  For light axions and thin walls, such that $m L \ll 1$, this transient regime can therefore span many orders of magnitude.

To be clear, if reflection off the wall occurs, then friction growing as $\gamma^2$ from reflecting particles is simple kinematics and not peculiar to axions. It is the fact that reflections can continue to occur for extremely relativistic speeds that is peculiar to light particles -- (pseudo-)Nambu-Goldstone bosons like axions and, equivalently, the longitudinal mode of a light but massive spin-one particle that changes mass across the wall~\cite{GarciaGarcia:2022yqb}. Of course axions are not the only dof that can receive corrections to their kinetic terms across an interface and, indeed, we should expect a small constant reflection probability for thin walls to be present more generally~\cite{prep}. However, axions are both naturally very light and produced cold, and so the regime of constant reflection is naturally active for an extended kinematic region. 

The general field theoretic treatment of section~\ref{sec:idea} allowed us to extend our results to regimes when a particle picture fails, by solving the coupled axion-wall equations of motion.  Particularly interesting is the result that even a light axion  $m \ll H$,  `frozen' during the time of the wall expansion, causes a frictional pressure which -- despite the suppression factor $m^2/H^2$ -- can stall the acceleration of relativistic walls. To our knowledge, Eq.\eqref{eq:P_intro} is the first example of friction acting even in the absence of any radiation or dark matter medium. This is particularly interesting for super-cooled phase transitions, where particle abundances are quickly diluted away, while energy density in a frozen axion field remains approximately constant \footnote{It may even grow in the case of the QCD axion, whose mass increases as temperature decreases.}. Even during inflation, walls could reach a constant subluminal expansion speed on subhorizon scales. In such `frozen' cases the energy released as the walls expand results in the appearance of an `axion cliff" -- a disturbance in the axion field that is sharply localized in front of the expanding wall.

Aside from their effect on the dynamics of expanding relativistic walls, the component of axion dark matter converted into relativistic particles can have an interesting and varied history long after, some aspects of which were discussed in this work. While bubble wall collisions rapidly dissipate energy, these relativistic axion shells may pass through each other, as discussed around Eq.\eqref{eq:ratio}. In certain scenarios, such reflected axions may remain relativistic long after, and even saturate the current bound on dark radiation $\Delta N_{\rm eff}\lesssim 0.3$ with a wide range of possible present day relativistic axion energies.  

In addition, we studied the effect of axions on the dynamics of the electroweak phase transition, which is often first order in extensions of the SM, and where nothing forbids the Higgs portal coupling of Eq.\eqref{eq:Lag_EFT}. Unsurprisingly, we find that the dominant source of friction is usually SM transition radiation, and for an axion to be more important requires significant super-cooling, pushing towards the lower limit of $T_*\sim \tilde{T}_c \sim 100 \MeV$. This, though, is not necessarily a fine-tuning, since in theories with classically scale invariant potentials $V(h,S)$  one can expect very strong cooling and the transition proceeds via QCD catalysis (see for example \cite{vonHarling:2017yew}). 

\paragraph{Outlook.}
Our goal in this work has been to highlight new effects rather than systematically study their phenomenological and observational consequences. We have not committed to an explicit model with a first order phase transition or domain wall collapse, but rather discussed different generic scenarios. We have also heavily focused on the misalignment mechanism in the harmonic approximation. 
It would be interesting to study fully calculable realisations in a few, motivated concrete models, and additionally map out the most likely range of energies of the relativistic axion background today, for a given decay constant. Moreover, while we have focused on the dimension-six coupling of Eq.\eqref{eq:dim6}, which cannot be forbidden by symmetry, it is worth considering the phenomenology of the more model-dependent dimension-five interaction in specific constructions. We showed that it is possible that most of the energy released in a phase transition can be deposited into ultra-relativistic axion `shells' and `cliffs' that, ignoring gravitational interactions, can pass through each other. Understanding their possible gravitational consequences and ultimate fate seems to us a particularly interesting goal.

\section*{Acknowledgments}
We wish to acknowledge and thank (now Dr.) Giacomo Koszegi for working with us on the early stages of this project. 
We are grateful to Anson Hook, Giovanni Villadoro, and Zach Weiner for useful discussions. The work of IGG is supported by the U.S.~Department of Energy grant No.~DE-SC0011637.

\color{black}

\appendix
\section{Appendix -- Regulating deltas} \label{app:deltareg}

In the main text, we derived the force per unit area acting on a thin bubble/domain wall, from a coherent background axion-like field with changing decay constant, and presented the final result
\be
    \mathcal{P} = - \frac{1}{A} \frac{\delta S_\text{axion}}{\delta z_0 (t)} = - {1\over 2} \Delta f^2 \int^\infty_{-\infty} dz' \delta\left(z'- z_0(t)\right) (\partial_\mu\theta )^2   = -\frac{1}{2} \Delta f^2 (\partial_\mu\theta )^2|_{\rm wall} \ .
    \label{eq:Cartoon}
\ee
However, in the case of an exact step function profile for the wall, the last expression is ill-defined, since $(\partial_\mu \theta)^2$ is discontinuous at the wall's location $z=z_0(t)$. This was stated explicitly in Eq.\eqref{eq:FchangeBCs} when deriving matching conditions.

It is tempting, but entirely wrong, to fix this by replacing the last line of Eq.\eqref{eq:Cartoon} with an average across the wall. If we regulated the delta function in the last integral as an appropriate limit of rectangles (or any other symmetric function, like a Gaussian) and kept $(\partial_\mu \theta)^2$ fixed, it would indeed lead us to an average. However, this delta function came from taking a derivative of the step function wall profile. Regulating means considering a different wall profile, which in turn affects $\theta$ through its EOM Eq.\eqref{eq:EOM_theta}. To take the limit correctly, one has to take into account both effects. This point is made beautifully through a simple example in a short article by Griffiths and Walborn \cite{Griffiths:1999deltas}.

To make sense of the expression in Eq.\eqref{eq:Cartoon} and check if the field theoretic derivation gives the same answer as the particle scattering result, we need to work a bit harder.
Consider regulating the wall profile with a linear interpolation of thickness $2\epsilon$:
\begin{align}
f_w^2 (t, z) &=
\begin{cases}
\ft^2 \ , &  z - z_0 < -\epsilon    \\ \ft^2-{\Delta f^2 \over 2 \epsilon } (z-z_0+\epsilon) \ , & -\epsilon <  z -z_0< \epsilon \\
f^2 \ , &  z -z_0 > \epsilon
\end{cases}
\label{eq:RegularisedLinearWall} \\
& = \ft^2 -  {\Delta f^2 \over 2 \epsilon}( z-z_0+\epsilon)  {\rm H}(z-z_0+\epsilon) 
+ {\Delta f^2 \over 2 \epsilon} \left( z - z_0 - \epsilon \right){\rm H}(z-z_0-\epsilon)
\label{eq:RegularisedWallThetaFunctions} 
\end{align}
where we take the centre of the wall to travel at constant speed $z_0(t) = v t$ as in the main text.
The step function is recovered in the limit $\epsilon\rightarrow 0$. Taking a derivative,
\begin{align}
    {\partial f^2( t,  z) \over \partial z_0(t)} = - { \Delta f^2 \over 2 \epsilon} \left\{ ( z-z_0-\epsilon)\delta(\hat z-z_0-\epsilon) + ( z-z_0+\epsilon)\delta( z-z_0+\epsilon) \right. \nonumber \\
   \left. - {\rm H}( z-z_0-\epsilon) + {\rm H}( z-z_0+\epsilon) \right\} \ . 
\end{align}
We can now evaluate pressure, towards which only the last two Heaviside functions contribute
\begin{align}
   \mathcal{P}(t) = -\frac{1}{A}{\delta S_\text{axion} \over \delta z_0(t)} = - \frac{1}{2}\int dz  {\partial f_w^2(t,z) \over \partial z_0} (\partial_\mu\theta_\epsilon )^2 =  -{\Delta f^2 \over 4 \epsilon}\int_{z_0(t)-\epsilon}^{z_0(t)+\epsilon} dz (\partial_\mu\theta_\epsilon )^2 \ , 
   \label{eq:regulatedPressure}
\end{align}
where we write $\theta_\epsilon$ to emphasize that $\theta$ also depends on $\epsilon$ through its EOM. This is the regulated version of Eq.\eqref{eq:Cartoon}. 
To evaluate Eq.\eqref{eq:regulatedPressure} now in the limit $\epsilon \rightarrow 0$, we need to specify the solution $\theta_{\epsilon}$. This will be a generalised version of Eq.\eqref{eq:theta_sol}, reducing to it in the limit. As in that case, we now work in the frame of the wall with appropriate coordinates 
$\hat t \equiv \gamma (t - vz) $ and $\hat z \equiv \gamma (z - vt)$. Its center is now located at $\hat z = 0$.
Letting $\theta_\epsilon(\hat t, \hat z) = e^{-i\omega \hat t}\chi(\hat z)$ as usual, the EOM becomes
\begin{align}
    -\omega^2 f_w^2(\hat z) \chi - \partial_{\hat z} \left( f_w^2(\hat z) \partial_{\hat z} \chi \right) = -m^2 f^2 \chi \ .
\end{align}
Since we are always interested in the relativistic regime, we simplify our life by now ignoring $m \rightarrow 0$.
There are three different domains to solve in, as per Eq.\eqref{eq:RegularisedLinearWall}. The solution with the scattering boundary conditions of interest is then
\begin{align}
\chi (\hat z) = \mathcal{N}
\begin{cases}
e^{-i \omega \hat z} + r_\theta  e^{ i \omega \hat z} \ , &   \hat z > \epsilon  \\ 
c_1 J_0\left[\left(\hat z+\epsilon +{2f^2 \over \Delta f^2} \epsilon \right)\omega\right] + c_2 Y_0\left[\left(\hat z+\epsilon +{2f^2 \over \Delta f^2} \epsilon \right)\omega\right]\ , & -\epsilon < \hat z < \epsilon \\
t_\theta e^{i\omega \hat z} \ , & \hat z < -\epsilon
\end{cases}
\label{eq:RegularisedSolution}
\end{align}
$J_n$ and $Y_n$ are Bessel functions of the first and second kind respectively, and $r_\theta,t_\theta,c_{1,2}$ are coefficients to be determined from the new appropriate matching conditions
\begin{align}
    \lim_{\delta \rightarrow 0}\left. \theta \right|_{\pm \epsilon+\delta}^{\pm \epsilon-\delta} = 0 \ , \qquad \lim_{\delta \rightarrow 0}\left. \partial_z\theta \right|_{\pm \epsilon+\delta}^{\pm \epsilon-\delta} = 0 ,
\end{align}
continuity of both $\theta$ and its derivative at the two domain boundaries $z=\pm \epsilon$. 
The explicit expressions for all the coefficients are easily obtained but are unnecessarily cumbersome to report here. 
They do of course have proper step function limit:
\begin{align}
    r_\theta \rightarrow \frac{f^2-\ft^2}{f^2+\ft^2} \ , \qquad 
    t_\theta \rightarrow \frac{2 f^2}{f^2+\ft^2} \ , \qquad \text{when} \quad \epsilon \rightarrow 0 \ ,
\end{align}
which are the same as Eq.(\ref{eq:RandT_f2}) in the massless limit.

As per Eq.\eqref{eq:regulatedPressure}, pressure is given by integrating over the region $-\epsilon < z < \epsilon$, precisely where the solution is non-trivial. Doing the integral explicitly is impossible, but we will still be able to extract the $\epsilon \rightarrow 0$ limit correctly.
Since $\theta_\epsilon$ is a real field we take the real part $\theta_{\epsilon} = {\rm R e}[e^{-i\omega \hat t}\chi(\hat z)]$, and proceed to evaluate Eq.\eqref{eq:regulatedPressure}. We focus on the $\hat z$ derivative part of $(\partial_\mu\theta_\epsilon)^2$, since herein lies the discontinuity in the step function limit in the rest frame. Rescaling this coordinate and then Taylor expanding,
\begin{align}
    \lim_{\epsilon\rightarrow 0} {1 \over 2 \epsilon} \int_{-\epsilon}^\epsilon d\hat z \,(\partial_{\hat z}\theta_{\epsilon})^2 & = \frac{1}{2}\lim_{\epsilon\rightarrow 0}\int_{-1}^1 d\bar{z} \,(\partial_{\bar{z}\epsilon}\theta_{\epsilon})^2  \ , \qquad \text{where}  \ \hat z = \bar{z} \epsilon \\
    & = \frac{1}{2}\lim_{\epsilon\rightarrow 0}\int_{-1}^1 d\bar{z} \, \left.(\partial_{\bar{z}\epsilon}\theta_{\epsilon})^2\right|_{\epsilon=0} + \mathcal{O}(\epsilon) \\
   &= -\mathcal{N}^2\int_{-1}^1 d\bar{z} \frac{8 \sin( \omega \hat t)^2 f^4 (f^2 + \Delta f^2)^2 \omega^2}{(2 f^2 + \Delta f^2)^2 \left(2 f^2 + (1 + \bar z) \Delta f^2\right)^2} \\
   &= \mathcal{N}^2\frac{4 f^2 \ft^2 \omega^2 \sin^2(\omega \hat t)}{(f^2 + \ft^2)^2} \ ,
   \label{eq:dzPart}
\end{align}
where in the third equality we showed the explicit leading order term in the $\epsilon$ expansion, given the solution Eq.\eqref{eq:RegularisedSolution}. Having got past the non-trivial limit we are now able to give a final answer:
\begin{align}
     \mathcal{P}(t)  &= - {\Delta f^2 \over 2}\lim_{\epsilon\rightarrow 0} {1 \over 2 \epsilon}\int_{-\epsilon}^\epsilon d\hat z \, (\partial_\mu \theta_{\epsilon} )^2 \\ 
     &= -{\Delta f^2 \over 2}\left[ \mathcal{N}^2 \omega^2 t_\theta^2 \sin^2(\omega \hat t) - \mathcal{N}^2\frac{4 f^2 \ft^2 \omega^2 \sin^2(\omega \hat t)}{(f^2 + \ft^2)^2} \right] \\
     &= 2\mathcal{N}^2  {f^2 r_\theta^2 \omega^2 } \sin^2(\omega \hat t) \ \rightarrow \  \mathcal{N}^2  {f^2 r_\theta^2 \omega^2 } \qquad \text{(averaging)}\\
     &= n_\theta r_\theta^2 2 \omega 
\end{align}
where we added the straightforward contribution proportional to $\dot{\theta}^2$, in the third line we perform the usual time average and in the final line we use the fact that incoming energy density is given by $\omega n_\theta = f^2 \mathcal{N}^2 \omega^2/2$, where $n$ is number density. The final expression
is exactly what we get in the particle picture from a flux of incoming massless particles $n$ reflecting with probability $\mathcal{R} = r_k^2$, as in Eq.(\ref{eq:ParticlePressure}).

We see that in this case getting the right answer was far easier in the particle picture. However, the field theoretic picture is more useful in other instances, such as the frozen axion field regime.

\section{Appendix -- Constraints on the axion-Higgs portal} \label{sec:Axion_Higgs_Portal_constraints}
We  estimate here cosmological and astrophysical bounds on the Axion-Higgs portal coupling, complementing the study of \cite{Bauer:2022rwf}.
\paragraph{Dark radiation:}
Given the relatively low bound Eq.\eqref{eq:Higgs_decay_bound}, one should check the thermal production of axions via this Higgs portal. Current bounds on dark radiation from Planck \cite{Aghanim:2018eyx} rule out a thermalized axion that decouples from the SM bath at temperatures $  T_{\rm dec.} \lesssim 100 \MeV$. Inserting the Higgs on vev we estimate the thermal production rate from fermion annihilation through an off-shell Higgs to be
\begin{align}
\label{eq:Rate_fermion_annihilation}
    \Gamma_{\rm th} \approx \frac{\kappa^2 m_f^2}{4\pi m_h^2 m_\rho^4} T^5
 \ , \quad \text{while} \quad m_f\lesssim T
\end{align}
Setting $\Gamma_{\rm th}=H$ we find the decoupling temperature to be roughly
\begin{align}
    T_{\rm dec.} \approx 1 \GeV \left( m_\rho / \sqrt{\eta} \over 10^{3} \GeV \right)^{4/3} \left( m_\mu \over m_f  \right)^{2/3} \ , \quad \text{while} \quad m_f\lesssim T_{\rm dec}
\end{align}
where we have normalised to the most relevant case of the muon. Given this rough estimate, we conclude that the present dark radiation bound is unlikely to be much stronger than Eq.\eqref{eq:Higgs_decay_bound}.

\paragraph{Supernovae:} light axions can be produced at the center of stars and escape; if this is too efficient, the star will lose enough energy that its dynamics would change appreciably. We make a very rough estimate of the bound for SN 1987A following the logic of \cite{Green:2021hjh}. The core of the collapsing supernova is hot enough to abundantly produce muons, which in turn will thermally produce axions via Eq.\eqref{eq:Rate_fermion_annihilation}. To avoid excessive cooling, this process should be inefficient over the relevant timescale, $\Gamma_{\rm th}^{-1}  \lesssim 1 s $. From this we obtain a bound of the same order of Eq.\eqref{eq:Higgs_decay_bound}. A more careful treatment of the complicated supernovae physics might therefore lead to the most stringent constraint on this axion-Higgs portal, but is beyond the scope of the current work.

\bibliography{main}

\providecommand{\href}[2]{#2}\begingroup\raggedright\begin{thebibliography}{10}

\bibitem{Bonati:2013tqa}
C.~Bonati, M.~D'Elia, P.~de~Forcrand, O.~Philipsen and F.~Sanfilippo,
  \emph{{The chiral phase transition for two-flavour QCD at imaginary and zero
  chemical potential}}, \href{http://dx.doi.org/10.22323/1.187.0219}{\emph{PoS}
  {\bf LATTICE2013} (2014) 219}, [\href{https://arxiv.org/abs/1311.0473}{{\tt
  1311.0473}}].

\bibitem{Borsanyi:2016ksw}
S.~Borsanyi et~al., \emph{{Calculation of the axion mass based on
  high-temperature lattice quantum chromodynamics}},
  \href{http://dx.doi.org/10.1038/nature20115}{\emph{Nature} {\bf 539} (2016)
  69--71}, [\href{https://arxiv.org/abs/1606.07494}{{\tt 1606.07494}}].

\bibitem{Kajantie:1996mn}
K.~Kajantie, M.~Laine, K.~Rummukainen and M.~E. Shaposhnikov, \emph{{Is there
  a~ hot electroweak phase transition at $m_H \gtrsim m_W$?}},
  \href{http://dx.doi.org/10.1103/PhysRevLett.77.2887}{\emph{Phys. Rev. Lett.}
  {\bf 77} (1996) 2887--2890},
  [\href{https://arxiv.org/abs/hep-ph/9605288}{{\tt hep-ph/9605288}}].

\bibitem{Kajantie:1996qd}
K.~Kajantie, M.~Laine, K.~Rummukainen and M.~E. Shaposhnikov, \emph{{A
  Nonperturbative analysis of the finite T phase transition in SU(2) x U(1)
  electroweak theory}},
  \href{http://dx.doi.org/10.1016/S0550-3213(97)00164-8}{\emph{Nucl. Phys. B}
  {\bf 493} (1997) 413--438},
  [\href{https://arxiv.org/abs/hep-lat/9612006}{{\tt hep-lat/9612006}}].

\bibitem{Laine:1998jb}
M.~Laine and K.~Rummukainen, \emph{{What's new with the electroweak phase
  transition?}},
  \href{http://dx.doi.org/10.1016/S0920-5632(99)85017-8}{\emph{Nucl. Phys. B
  Proc. Suppl.} {\bf 73} (1999) 180--185},
  [\href{https://arxiv.org/abs/hep-lat/9809045}{{\tt hep-lat/9809045}}].

\bibitem{Csikor:1998eu}
F.~Csikor, Z.~Fodor and J.~Heitger, \emph{{Endpoint of the hot electroweak
  phase transition}},
  \href{http://dx.doi.org/10.1103/PhysRevLett.82.21}{\emph{Phys. Rev. Lett.}
  {\bf 82} (1999) 21--24}, [\href{https://arxiv.org/abs/hep-ph/9809291}{{\tt
  hep-ph/9809291}}].

\bibitem{Caprini:2015zlo}
C.~Caprini et~al., \emph{{Science with the space-based interferometer eLISA.
  II: Gravitational waves from cosmological phase transitions}},
  \href{http://dx.doi.org/10.1088/1475-7516/2016/04/001}{\emph{JCAP} {\bf 04}
  (2016) 001}, [\href{https://arxiv.org/abs/1512.06239}{{\tt 1512.06239}}].

\bibitem{Caprini:2019egz}
C.~Caprini et~al., \emph{{Detecting gravitational waves from cosmological phase
  transitions with LISA: an update}},
  \href{http://dx.doi.org/10.1088/1475-7516/2020/03/024}{\emph{JCAP} {\bf 03}
  (2020) 024}, [\href{https://arxiv.org/abs/1910.13125}{{\tt 1910.13125}}].

\bibitem{Peccei:1977hh}
R.~D. Peccei and H.~R. Quinn, \emph{{CP Conservation in the Presence of
  Instantons}},
  \href{http://dx.doi.org/10.1103/PhysRevLett.38.1440}{\emph{Phys. Rev. Lett.}
  {\bf 38} (1977) 1440--1443}.

\bibitem{Weinberg:1977ma}
S.~Weinberg, \emph{{A New Light Boson?}},
  \href{http://dx.doi.org/10.1103/PhysRevLett.40.223}{\emph{Phys. Rev. Lett.}
  {\bf 40} (1978) 223--226}.

\bibitem{Wilczek:1977pj}
F.~Wilczek, \emph{{Problem of Strong $P$ and $T$ Invariance in the Presence of
  Instantons}}, \href{http://dx.doi.org/10.1103/PhysRevLett.40.279}{\emph{Phys.
  Rev. Lett.} {\bf 40} (1978) 279--282}.

\bibitem{Arvanitaki:2009fg}
A.~Arvanitaki, S.~Dimopoulos, S.~Dubovsky, N.~Kaloper and J.~March-Russell,
  \emph{{String Axiverse}},
  \href{http://dx.doi.org/10.1103/PhysRevD.81.123530}{\emph{Phys. Rev. D} {\bf
  81} (2010) 123530}, [\href{https://arxiv.org/abs/0905.4720}{{\tt
  0905.4720}}].

\bibitem{Sikivie:1982qv}
P.~Sikivie, \emph{{Of Axions, Domain Walls and the Early Universe}},
  \href{http://dx.doi.org/10.1103/PhysRevLett.48.1156}{\emph{Phys. Rev. Lett.}
  {\bf 48} (1982) 1156--1159}.

\bibitem{Vilenkin:1982ks}
A.~Vilenkin and A.~E. Everett, \emph{{Cosmic Strings and Domain Walls in Models
  with Goldstone and PseudoGoldstone Bosons}},
  \href{http://dx.doi.org/10.1103/PhysRevLett.48.1867}{\emph{Phys. Rev. Lett.}
  {\bf 48} (1982) 1867--1870}.

\bibitem{Vilenkin:1984ib}
A.~Vilenkin, \emph{{Cosmic Strings and Domain Walls}},
  \href{http://dx.doi.org/10.1016/0370-1573(85)90033-X}{\emph{Phys. Rept.} {\bf
  121} (1985) 263--315}.

\bibitem{Davis:1986xc}
R.~L. Davis, \emph{{Cosmic Axions from Cosmic Strings}},
  \href{http://dx.doi.org/10.1016/0370-2693(86)90300-X}{\emph{Phys. Lett. B}
  {\bf 180} (1986) 225--230}.

\bibitem{Gorghetto:2018myk}
M.~Gorghetto, E.~Hardy and G.~Villadoro, \emph{{Axions from Strings: the
  Attractive Solution}},
  \href{http://dx.doi.org/10.1007/JHEP07(2018)151}{\emph{JHEP} {\bf 07} (2018)
  151}, [\href{https://arxiv.org/abs/1806.04677}{{\tt 1806.04677}}].

\bibitem{Preskill:1982cy}
J.~Preskill, M.~B. Wise and F.~Wilczek, \emph{{Cosmology of the Invisible
  Axion}}, \href{http://dx.doi.org/10.1016/0370-2693(83)90637-8}{\emph{Phys.
  Lett. B} {\bf 120} (1983) 127--132}.

\bibitem{Abbott:1982af}
L.~F. Abbott and P.~Sikivie, \emph{{A Cosmological Bound on the Invisible
  Axion}}, \href{http://dx.doi.org/10.1016/0370-2693(83)90638-X}{\emph{Phys.
  Lett. B} {\bf 120} (1983) 133--136}.

\bibitem{Dine:1982ah}
M.~Dine and W.~Fischler, \emph{{The Not So Harmless Axion}},
  \href{http://dx.doi.org/10.1016/0370-2693(83)90639-1}{\emph{Phys. Lett. B}
  {\bf 120} (1983) 137--141}.

\bibitem{Zeldovich:1974uw}
Y.~B. Zeldovich, I.~Y. Kobzarev and L.~B. Okun, \emph{{Cosmological
  Consequences of the Spontaneous Breakdown of Discrete Symmetry}}, {\emph{Zh.
  Eksp. Teor. Fiz.} {\bf 67} (1974) 3--11}.

\bibitem{Vachaspati:2017hjw}
T.~Vachaspati, \emph{{Lunar Mass Black Holes from QCD Axion Cosmology}},
  \href{https://arxiv.org/abs/1706.03868}{{\tt 1706.03868}}.

\bibitem{Ferrer:2018uiu}
F.~Ferrer, E.~Masso, G.~Panico, O.~Pujolas and F.~Rompineve, \emph{{Primordial
  Black Holes from the QCD axion}},
  \href{http://dx.doi.org/10.1103/PhysRevLett.122.101301}{\emph{Phys. Rev.
  Lett.} {\bf 122} (2019) 101301},
  [\href{https://arxiv.org/abs/1807.01707}{{\tt 1807.01707}}].

\bibitem{Gelmini:2022nim}
G.~B. Gelmini, A.~Simpson and E.~Vitagliano, \emph{{Catastrogenesis: DM, GWs,
  and PBHs from ALP string-wall networks}},
  \href{http://dx.doi.org/10.1088/1475-7516/2023/02/031}{\emph{JCAP} {\bf 02}
  (2023) 031}, [\href{https://arxiv.org/abs/2207.07126}{{\tt 2207.07126}}].

\bibitem{Gouttenoire:2023gbn}
Y.~Gouttenoire and E.~Vitagliano, \emph{{Primordial black holes and wormholes
  from domain wall networks}},
  \href{http://dx.doi.org/10.1103/PhysRevD.109.123507}{\emph{Phys. Rev. D} {\bf
  109} (2024) 123507}, [\href{https://arxiv.org/abs/2311.07670}{{\tt
  2311.07670}}].

\bibitem{Ferreira:2024eru}
R.~Z. Ferreira, A.~Notari, O.~Pujol\`as and F.~Rompineve, \emph{{Collapsing
  domain wall networks: impact on pulsar timing arrays and primordial black
  holes}}, \href{http://dx.doi.org/10.1088/1475-7516/2024/06/020}{\emph{JCAP}
  {\bf 06} (2024) 020}, [\href{https://arxiv.org/abs/2401.14331}{{\tt
  2401.14331}}].

\bibitem{Dunsky:2024zdo}
D.~I. Dunsky and M.~Kongsore, \emph{{Primordial Black Holes from Axion Domain
  Wall Collapse}},  \href{https://arxiv.org/abs/2402.03426}{{\tt 2402.03426}}.

\bibitem{GarciaGarcia:2022yqb}
I.~Garcia~Garcia, G.~Koszegi and R.~Petrossian-Byrne, \emph{{Reflections on
  bubble walls}}, \href{http://dx.doi.org/10.1007/JHEP09(2023)013}{\emph{JHEP}
  {\bf 09} (2023) 013}, [\href{https://arxiv.org/abs/2212.10572}{{\tt
  2212.10572}}].

\bibitem{Turok:1992jp}
N.~Turok, \emph{{Electroweak bubbles: Nucleation and growth}},
  \href{http://dx.doi.org/10.1103/PhysRevLett.68.1803}{\emph{Phys. Rev. Lett.}
  {\bf 68} (1992) 1803--1806}.

\bibitem{Dine:1992wr}
M.~Dine, R.~G. Leigh, P.~Y. Huet, A.~D. Linde and D.~A. Linde, \emph{{Towards
  the theory of the electroweak phase transition}},
  \href{http://dx.doi.org/10.1103/PhysRevD.46.550}{\emph{Phys. Rev. D} {\bf 46}
  (1992) 550--571}, [\href{https://arxiv.org/abs/hep-ph/9203203}{{\tt
  hep-ph/9203203}}].

\bibitem{Liu:1992tn}
B.-H. Liu, L.~D. McLerran and N.~Turok, \emph{{Bubble nucleation and growth at
  a baryon number producing electroweak phase transition}},
  \href{http://dx.doi.org/10.1103/PhysRevD.46.2668}{\emph{Phys. Rev. D} {\bf
  46} (1992) 2668--2688}.

\bibitem{Arnold:1993wc}
P.~B. Arnold, \emph{{One loop fluctuation - dissipation formula for bubble wall
  velocity}}, \href{http://dx.doi.org/10.1103/PhysRevD.48.1539}{\emph{Phys.
  Rev. D} {\bf 48} (1993) 1539--1545},
  [\href{https://arxiv.org/abs/hep-ph/9302258}{{\tt hep-ph/9302258}}].

\bibitem{Moore:1995ua}
G.~D. Moore and T.~Prokopec, \emph{{Bubble wall velocity in a first order
  electroweak phase transition}},
  \href{http://dx.doi.org/10.1103/PhysRevLett.75.777}{\emph{Phys. Rev. Lett.}
  {\bf 75} (1995) 777--780}, [\href{https://arxiv.org/abs/hep-ph/9503296}{{\tt
  hep-ph/9503296}}].

\bibitem{Moore:1995si}
G.~D. Moore and T.~Prokopec, \emph{{How fast can the wall move? A Study of the
  electroweak phase transition dynamics}},
  \href{http://dx.doi.org/10.1103/PhysRevD.52.7182}{\emph{Phys. Rev. D} {\bf
  52} (1995) 7182--7204}, [\href{https://arxiv.org/abs/hep-ph/9506475}{{\tt
  hep-ph/9506475}}].

\bibitem{Bodeker:2009qy}
D.~Bodeker and G.~D. Moore, \emph{{Can electroweak bubble walls run away?}},
  \href{http://dx.doi.org/10.1088/1475-7516/2009/05/009}{\emph{JCAP} {\bf 05}
  (2009) 009}, [\href{https://arxiv.org/abs/0903.4099}{{\tt 0903.4099}}].

\bibitem{Espinosa:2010hh}
J.~R. Espinosa, T.~Konstandin, J.~M. No and G.~Servant, \emph{{Energy Budget of
  Cosmological First-order Phase Transitions}},
  \href{http://dx.doi.org/10.1088/1475-7516/2010/06/028}{\emph{JCAP} {\bf 06}
  (2010) 028}, [\href{https://arxiv.org/abs/1004.4187}{{\tt 1004.4187}}].

\bibitem{Leitao:2015ola}
L.~Leitao and A.~Megevand, \emph{{Hydrodynamics of ultra-relativistic bubble
  walls}}, \href{http://dx.doi.org/10.1016/j.nuclphysb.2016.02.009}{\emph{Nucl.
  Phys. B} {\bf 905} (2016) 45--72},
  [\href{https://arxiv.org/abs/1510.07747}{{\tt 1510.07747}}].

\bibitem{Bodeker:2017cim}
D.~Bodeker and G.~D. Moore, \emph{{Electroweak Bubble Wall Speed Limit}},
  \href{http://dx.doi.org/10.1088/1475-7516/2017/05/025}{\emph{JCAP} {\bf 05}
  (2017) 025}, [\href{https://arxiv.org/abs/1703.08215}{{\tt 1703.08215}}].

\bibitem{Azatov:2020ufh}
A.~Azatov and M.~Vanvlasselaer, \emph{{Bubble wall velocity: heavy physics
  effects}}, \href{http://dx.doi.org/10.1088/1475-7516/2021/01/058}{\emph{JCAP}
  {\bf 01} (2021) 058}, [\href{https://arxiv.org/abs/2010.02590}{{\tt
  2010.02590}}].

\bibitem{Ai:2021kak}
W.-Y. Ai, B.~Garbrecht and C.~Tamarit, \emph{{Bubble wall velocities in local
  equilibrium}},
  \href{http://dx.doi.org/10.1088/1475-7516/2022/03/015}{\emph{JCAP} {\bf 03}
  (2022) 015}, [\href{https://arxiv.org/abs/2109.13710}{{\tt 2109.13710}}].

\bibitem{Mancha_2021}
M.~B. Mancha, T.~Prokopec and B.~{\'{S}}wie{\.{z}}ewska, \emph{Field-theoretic
  derivation of bubble-wall force},
  \href{http://dx.doi.org/10.1007/jhep01(2021)070}{\emph{Journal of High Energy
  Physics} {\bf 2021} (jan, 2021) }.

\bibitem{Bigazzi:2021ucw}
F.~Bigazzi, A.~Caddeo, T.~Canneti and A.~L. Cotrone, \emph{{Bubble wall
  velocity at strong coupling}},
  \href{http://dx.doi.org/10.1007/JHEP08(2021)090}{\emph{JHEP} {\bf 08} (2021)
  090}, [\href{https://arxiv.org/abs/2104.12817}{{\tt 2104.12817}}].

\bibitem{Bea:2021zsu}
Y.~Bea, J.~Casalderrey-Solana, T.~Giannakopoulos, D.~Mateos,
  M.~Sanchez-Garitaonandia and M.~Zilh\~ao, \emph{{Bubble wall velocity from
  holography}},
  \href{http://dx.doi.org/10.1103/PhysRevD.104.L121903}{\emph{Phys. Rev. D}
  {\bf 104} (2021) L121903}, [\href{https://arxiv.org/abs/2104.05708}{{\tt
  2104.05708}}].

\bibitem{Gouttenoire:2021kjv}
Y.~Gouttenoire, R.~Jinno and F.~Sala, \emph{{Friction pressure on relativistic
  bubble walls}}, \href{http://dx.doi.org/10.1007/JHEP05(2022)004}{\emph{JHEP}
  {\bf 05} (2022) 004}, [\href{https://arxiv.org/abs/2112.07686}{{\tt
  2112.07686}}].

\bibitem{Azatov:2022tii}
A.~Azatov, G.~Barni, S.~Chakraborty, M.~Vanvlasselaer and W.~Yin,
  \emph{{Ultra-relativistic bubbles from the simplest Higgs portal and their
  cosmological consequences}},
  \href{http://dx.doi.org/10.1007/JHEP10(2022)017}{\emph{JHEP} {\bf 10} (2022)
  017}, [\href{https://arxiv.org/abs/2207.02230}{{\tt 2207.02230}}].

\bibitem{Laurent:2022jrs}
B.~Laurent and J.~M. Cline, \emph{{First principles determination of bubble
  wall velocity}},
  \href{http://dx.doi.org/10.1103/PhysRevD.106.023501}{\emph{Phys. Rev. D} {\bf
  106} (2022) 023501}, [\href{https://arxiv.org/abs/2204.13120}{{\tt
  2204.13120}}].

\bibitem{DeCurtis:2022hlx}
S.~De~Curtis, L.~D. Rose, A.~Guiggiani, A.~G. Muyor and G.~Panico,
  \emph{{Bubble wall dynamics at the electroweak phase transition}},
  \href{http://dx.doi.org/10.1007/JHEP03(2022)163}{\emph{JHEP} {\bf 03} (2022)
  163}, [\href{https://arxiv.org/abs/2201.08220}{{\tt 2201.08220}}].

\bibitem{Wang:2022txy}
S.-J. Wang and Z.-Y. Yuwen, \emph{{Hydrodynamic backreaction force of
  cosmological bubble expansion}},
  \href{http://dx.doi.org/10.1103/PhysRevD.107.023501}{\emph{Phys. Rev. D} {\bf
  107} (2023) 023501}, [\href{https://arxiv.org/abs/2205.02492}{{\tt
  2205.02492}}].

\bibitem{Coleman:1977py}
S.~R. Coleman, \emph{{The Fate of the False Vacuum. 1. Semiclassical Theory}},
  \href{http://dx.doi.org/10.1103/PhysRevD.16.1248}{\emph{Phys. Rev. D} {\bf
  15} (1977) 2929--2936}.

\bibitem{GrillidiCortona:2015jxo}
G.~Grilli~di Cortona, E.~Hardy, J.~Pardo~Vega and G.~Villadoro, \emph{{The QCD
  axion, precisely}},
  \href{http://dx.doi.org/10.1007/JHEP01(2016)034}{\emph{JHEP} {\bf 01} (2016)
  034}, [\href{https://arxiv.org/abs/1511.02867}{{\tt 1511.02867}}].

\bibitem{AxionLimits}
C.~O'Hare, ``cajohare/axionlimits: Axionlimits.''
  \url{https://cajohare.github.io/AxionLimits/}, July, 2020.
\newblock 10.5281/zenodo.3932430.

\bibitem{Kosowsky:1991ua}
A.~Kosowsky, M.~S. Turner and R.~Watkins, \emph{{Gravitational radiation from
  colliding vacuum bubbles}},
  \href{http://dx.doi.org/10.1103/PhysRevD.45.4514}{\emph{Phys. Rev. D} {\bf
  45} (1992) 4514--4535}.

\bibitem{Kosowsky:1992rz}
A.~Kosowsky, M.~S. Turner and R.~Watkins, \emph{{Gravitational waves from first
  order cosmological phase transitions}},
  \href{http://dx.doi.org/10.1103/PhysRevLett.69.2026}{\emph{Phys. Rev. Lett.}
  {\bf 69} (1992) 2026--2029}.

\bibitem{Kosowsky:1992vn}
A.~Kosowsky and M.~S. Turner, \emph{{Gravitational radiation from colliding
  vacuum bubbles: envelope approximation to many bubble collisions}},
  \href{http://dx.doi.org/10.1103/PhysRevD.47.4372}{\emph{Phys. Rev. D} {\bf
  47} (1993) 4372--4391}, [\href{https://arxiv.org/abs/astro-ph/9211004}{{\tt
  astro-ph/9211004}}].

\bibitem{Kamionkowski:1993fg}
M.~Kamionkowski, A.~Kosowsky and M.~S. Turner, \emph{{Gravitational radiation
  from first order phase transitions}},
  \href{http://dx.doi.org/10.1103/PhysRevD.49.2837}{\emph{Phys. Rev. D} {\bf
  49} (1994) 2837--2851}, [\href{https://arxiv.org/abs/astro-ph/9310044}{{\tt
  astro-ph/9310044}}].

\bibitem{Hook:2020phx}
A.~Hook, G.~Marques-Tavares and D.~Racco, \emph{{Causal gravitational waves as
  a probe of free streaming particles and the expansion of the Universe}},
  \href{http://dx.doi.org/10.1007/JHEP02(2021)117}{\emph{JHEP} {\bf 02} (2021)
  117}, [\href{https://arxiv.org/abs/2010.03568}{{\tt 2010.03568}}].

\bibitem{Loverde:2022wih}
M.~Loverde and Z.~J. Weiner, \emph{{Probing neutrino interactions and dark
  radiation with gravitational waves}},
  \href{http://dx.doi.org/10.1088/1475-7516/2023/02/064}{\emph{JCAP} {\bf 02}
  (2023) 064}, [\href{https://arxiv.org/abs/2208.11714}{{\tt 2208.11714}}].

\bibitem{Aghanim:2018eyx}
{\scshape Planck} collaboration, N.~Aghanim et~al., \emph{{Planck 2018 results.
  VI. Cosmological parameters}},
  \href{http://dx.doi.org/10.1051/0004-6361/201833910}{\emph{Astron.
  Astrophys.} {\bf 641} (2020) A6},
  [\href{https://arxiv.org/abs/1807.06209}{{\tt 1807.06209}}].

\bibitem{Raffelt:1996wa}
G.~G. Raffelt, \emph{{Stars as laboratories for fundamental physics}: {The
  astrophysics of neutrinos, axions, and other weakly interacting particles}}.
\newblock 5, 1996.

\bibitem{CAST:2015qbl}
{\scshape CAST} collaboration, M.~Arik et~al., \emph{{New solar axion search
  using the CERN Axion Solar Telescope with $^4$He filling}},
  \href{http://dx.doi.org/10.1103/PhysRevD.92.021101}{\emph{Phys. Rev. D} {\bf
  92} (2015) 021101}, [\href{https://arxiv.org/abs/1503.00610}{{\tt
  1503.00610}}].

\bibitem{IAXO:2019mpb}
{\scshape IAXO} collaboration, E.~Armengaud et~al., \emph{{Physics potential of
  the International Axion Observatory (IAXO)}},
  \href{http://dx.doi.org/10.1088/1475-7516/2019/06/047}{\emph{JCAP} {\bf 06}
  (2019) 047}, [\href{https://arxiv.org/abs/1904.09155}{{\tt 1904.09155}}].

\bibitem{IAXO:2020wwp}
{\scshape IAXO} collaboration, A.~Abeln et~al., \emph{{Conceptual design of
  BabyIAXO, the intermediate stage towards the International Axion
  Observatory}}, \href{http://dx.doi.org/10.1007/JHEP05(2021)137}{\emph{JHEP}
  {\bf 05} (2021) 137}, [\href{https://arxiv.org/abs/2010.12076}{{\tt
  2010.12076}}].

\bibitem{2017arXiv170200786A}
P.~{Amaro-Seoane}, H.~{Audley}, S.~{Babak}, J.~{Baker}, E.~{Barausse},
  P.~{Bender} et~al., \emph{{Laser Interferometer Space Antenna}},
  \href{http://dx.doi.org/10.48550/arXiv.1702.00786}{\emph{arXiv e-prints}
  (Feb., 2017) arXiv:1702.00786}, [\href{https://arxiv.org/abs/1702.00786}{{\tt
  1702.00786}}].

\bibitem{LISACosmologyWorkingGroup:2022jok}
{\scshape LISA Cosmology Working Group} collaboration, P.~Auclair et~al.,
  \emph{{Cosmology with the Laser Interferometer Space Antenna}},
  \href{http://dx.doi.org/10.1007/s41114-023-00045-2}{\emph{Living Rev. Rel.}
  {\bf 26} (2023) 5}, [\href{https://arxiv.org/abs/2204.05434}{{\tt
  2204.05434}}].

\bibitem{Kuzmin:1985mm}
V.~A. Kuzmin, V.~A. Rubakov and M.~E. Shaposhnikov, \emph{{On the Anomalous
  Electroweak Baryon Number Nonconservation in the Early Universe}},
  \href{http://dx.doi.org/10.1016/0370-2693(85)91028-7}{\emph{Phys. Lett. B}
  {\bf 155} (1985) 36}.

\bibitem{Shaposhnikov:1986jp}
M.~E. Shaposhnikov, \emph{{Possible Appearance of the Baryon Asymmetry of the
  Universe in an Electroweak Theory}}, {\emph{JETP Lett.} {\bf 44} (1986)
  465--468}.

\bibitem{Shaposhnikov:1987tw}
M.~E. Shaposhnikov, \emph{{Baryon Asymmetry of the Universe in Standard
  Electroweak Theory}},
  \href{http://dx.doi.org/10.1016/0550-3213(87)90127-1}{\emph{Nucl. Phys. B}
  {\bf 287} (1987) 757--775}.

\bibitem{Morrissey:2012db}
D.~E. Morrissey and M.~J. Ramsey-Musolf, \emph{{Electroweak baryogenesis}},
  \href{http://dx.doi.org/10.1088/1367-2630/14/12/125003}{\emph{New J. Phys.}
  {\bf 14} (2012) 125003}, [\href{https://arxiv.org/abs/1206.2942}{{\tt
  1206.2942}}].

\bibitem{Creminelli:2001th}
P.~Creminelli, A.~Nicolis and R.~Rattazzi, \emph{{Holography and the
  electroweak phase transition}},
  \href{http://dx.doi.org/10.1088/1126-6708/2002/03/051}{\emph{JHEP} {\bf 03}
  (2002) 051}, [\href{https://arxiv.org/abs/hep-th/0107141}{{\tt
  hep-th/0107141}}].

\bibitem{Craig:2014lda}
N.~Craig, H.~K. Lou, M.~McCullough and A.~Thalapillil, \emph{{The Higgs Portal
  Above Threshold}},
  \href{http://dx.doi.org/10.1007/JHEP02(2016)127}{\emph{JHEP} {\bf 02} (2016)
  127}, [\href{https://arxiv.org/abs/1412.0258}{{\tt 1412.0258}}].

\bibitem{Witten:1984rs}
E.~Witten, \emph{{Cosmic Separation of Phases}},
  \href{http://dx.doi.org/10.1103/PhysRevD.30.272}{\emph{Phys. Rev. D} {\bf 30}
  (1984) 272--285}.

\bibitem{Vanvlasselaer:2020niz}
A.~Azatov and M.~Vanvlasselaer, \emph{{Bubble wall velocity: heavy physics
  effects}}, \href{http://dx.doi.org/10.1088/1475-7516/2021/01/058}{\emph{JCAP}
  {\bf 01} (2021) 058}, [\href{https://arxiv.org/abs/2010.02590}{{\tt
  2010.02590}}].

\bibitem{Azatov:2023xem}
A.~Azatov, G.~Barni, R.~Petrossian-Byrne and M.~Vanvlasselaer,
  \emph{{Quantisation Across Bubble Walls and Friction}},
  \href{https://arxiv.org/abs/2310.06972}{{\tt 2310.06972}}.

\bibitem{Profumo:2007wc}
S.~Profumo, M.~J. Ramsey-Musolf and G.~Shaughnessy, \emph{{Singlet Higgs
  phenomenology and the electroweak phase transition}},
  \href{http://dx.doi.org/10.1088/1126-6708/2007/08/010}{\emph{JHEP} {\bf 08}
  (2007) 010}, [\href{https://arxiv.org/abs/0705.2425}{{\tt 0705.2425}}].

\bibitem{Espinosa:2011ax}
J.~R. Espinosa, T.~Konstandin and F.~Riva, \emph{{Strong Electroweak Phase
  Transitions in the Standard Model with a Singlet}},
  \href{http://dx.doi.org/10.1016/j.nuclphysb.2011.09.010}{\emph{Nucl. Phys. B}
  {\bf 854} (2012) 592--630}, [\href{https://arxiv.org/abs/1107.5441}{{\tt
  1107.5441}}].

\bibitem{Bauer:2022rwf}
M.~Bauer, G.~Rostagni and J.~Spinner, \emph{{Axion-Higgs portal}},
  \href{http://dx.doi.org/10.1103/PhysRevD.107.015007}{\emph{Phys. Rev. D} {\bf
  107} (2023) 015007}, [\href{https://arxiv.org/abs/2207.05762}{{\tt
  2207.05762}}].

\bibitem{prep}
I.~Garcia~Garcia, A.~Hook and R.~Petrossian-Byrne, \emph{{in preparation}}.

\bibitem{vonHarling:2017yew}
B.~von Harling and G.~Servant, \emph{{QCD-induced Electroweak Phase
  Transition}}, \href{http://dx.doi.org/10.1007/JHEP01(2018)159}{\emph{JHEP}
  {\bf 01} (2018) 159}, [\href{https://arxiv.org/abs/1711.11554}{{\tt
  1711.11554}}].

\bibitem{Griffiths:1999deltas}
D.~Griffiths and S.~Walborn, \emph{{Dirac deltas and discontinuous functions}},
  \href{http://dx.doi.org/10.1119/1.19283}{\emph{American Journal of Physics}
  {\bf 67} (05, 1999) 446--447},
  [\href{https://arxiv.org/abs/https://pubs.aip.org/aapt/ajp/article-pdf/67/5/446/10115861/446\_1\_online.pdf}{{\tt
  https://pubs.aip.org/aapt/ajp/article-pdf/67/5/446/10115861/446\_1\_online.pdf}}].

\bibitem{Green:2021hjh}
D.~Green, Y.~Guo and B.~Wallisch, \emph{{Cosmological implications of
  axion-matter couplings}},
  \href{http://dx.doi.org/10.1088/1475-7516/2022/02/019}{\emph{JCAP} {\bf 02}
  (2022) 019}, [\href{https://arxiv.org/abs/2109.12088}{{\tt 2109.12088}}].

\end{thebibliography}\endgroup


\providecommand{\href}[2]{#2}\begingroup\raggedright\begin{thebibliography}{1}

\bibitem{Elor:2023xbz}
G.~Elor, R.~Jinno, S.~Kumar, R.~McGehee and Y.~Tsai, \emph{{Finite Bubble
  Statistics Constrain Late Cosmological Phase Transitions}},
  \href{https://arxiv.org/abs/2311.16222}{{\tt 2311.16222}}.

\bibitem{Laine:1998jb}
M.~Laine and K.~Rummukainen, \emph{{What's new with the electroweak phase
  transition?}},
  \href{http://dx.doi.org/10.1016/S0920-5632(99)85017-8}{\emph{Nucl. Phys. B
  Proc. Suppl.} {\bf 73} (1999) 180--185},
  [\href{https://arxiv.org/abs/hep-lat/9809045}{{\tt hep-lat/9809045}}].

\bibitem{Ellis:2019oqb}
J.~Ellis, M.~Lewicki, J.~M. No and V.~Vaskonen, \emph{{Gravitational wave
  energy budget in strongly supercooled phase transitions}},
  \href{http://dx.doi.org/10.1088/1475-7516/2019/06/024}{\emph{JCAP} {\bf 06}
  (2019) 024}, [\href{https://arxiv.org/abs/1903.09642}{{\tt 1903.09642}}].

\bibitem{Azatov:2022tii}
A.~Azatov, G.~Barni, S.~Chakraborty, M.~Vanvlasselaer and W.~Yin,
  \emph{{Ultra-relativistic bubbles from the simplest Higgs portal and their
  cosmological consequences}},
  \href{http://dx.doi.org/10.1007/JHEP10(2022)017}{\emph{JHEP} {\bf 10} (2022)
  017}, [\href{https://arxiv.org/abs/2207.02230}{{\tt 2207.02230}}].

\end{thebibliography}\endgroup

\end{document}